\documentclass[11pt]{article}
\usepackage{hyperref}
\usepackage{amssymb,amsmath,amsfonts,mathtools}
\usepackage{epsfig}

\usepackage[sort,compress]{cite}
\usepackage[bulletsep]{collref}

\def\clock{{\count0=\time
           \divide\count0 60
           \ifnum\count0<10 0\fi\the\count0
           \multiply\count0 -60 \advance\count0 \time
           :\ifnum\count0<10 0\fi \the\count0
         }}
\newcommand{\timestamp}{{\small\vbox{\hbox{\tt\jobname.tex}
\hbox{\the\day/\the\month/\the\year, \clock}}}}

\newcommand{\CL}{\mathcal{L}}

\newcommand{\CP}{\mathcal{P}}

\newcommand{\C}{\mathbb{C}}

\newcommand{\nn}{\nonumber}

\DeclareMathOperator{\tr}{Tr}

\DeclareMathOperator{\vol}{vol}

\newcommand{\ads}{\textup{\textrm{AdS}}}
\newcommand{\cft}{\textup{\textrm{CFT}}}
\newcommand{\sphere}{\textup{\textrm{S}}}
\newcommand{\torus}{\textup{\textrm{T}}}

\newcommand{\cp}{\C \textup{\textrm{P}}}

\newcommand{\order}{\mathcal{O}}

\makeatletter
\let\old@startsection=\@startsection
\let\oldl@section=\l@section
\renewcommand{\@startsection}[6]{\old@startsection{#1}{#2}{#3}{#4}{#5}{#6\mathversion{bold}}}
\renewcommand{\l@section}[2]{\oldl@section{\mathversion{bold}#1}{#2}}
\makeatother

\numberwithin{equation}{section}

\def\[{\begin{equation}}
\def\]{\end{equation}}
\newcommand{\be}{\begin{eqnarray}}
\newcommand{\ee}{\end{eqnarray}}

\newcommand{\alg}[1]{\mathfrak{#1}}
\newcommand{\grp}[1]{\mathrm{#1}}

\newcommand{\grSU}{\grp{SU}}

\newcommand{\grSL}{\grp{SL}}

\newcommand{\algOSp}{\alg{osp}}

\newcommand{\algPSU}{\alg{psu}}

\newcommand{\quarter}{\frac{1}{4}}
\newcommand{\half}{\frac{1}{2}}
\newcommand{\p}{\partial}

\usepackage{color}

\def \EE {{\mathbb{E}}}
\def \KK {{\mathbb{K}}}

\def\no{\nonumber}

\def\S{\mathcal{S}}
\def\r{\rho}
\def\a{\alpha}
\def\k{\kappa}
\def\E{\mathcal{E}}

\DeclareMathOperator{\sn}{sn}
\DeclareMathOperator{\cn}{cn}
\DeclareMathOperator{\dn}{dn}

\def\s{\sigma}
\def\d{\partial}

\def\r{\rho}
\def\O{\mathcal{O}}

\def\o{\omega}

\def \Om {\Omega }
\def\td{\tilde}

\begin{document}
\renewcommand{\thefootnote}{\arabic{footnote}}

\overfullrule=0pt
\parskip=2pt
\parindent=12pt
\headheight=0in \headsep=0in \topmargin=0in \oddsidemargin=0in

\vspace{ -3cm} \thispagestyle{empty} \vspace{-1cm}
\begin{flushright} 
\footnotesize
ICCUB-12-127 \\
HU-EP-12/11\\
NORDITA-2012-22 \\
ITP-UU-12/11 \\
SPIN-12/10
\end{flushright}%

\begin{center}
\vspace{1.2cm}
{\Large\bf \mathversion{bold}
Generalized cusp in  $\ads_4\times \cp^3$ \\
 \vspace{0.2cm}
and more one-loop results from semiclassical strings 
}

 \vspace{0.8cm} {
  V.~Forini$^{a,b,}$\footnote{vforini@icc.ub.edu} ,
 V.~Giangreco M. Puletti$^{c,}$\footnote{marotta@chalmers.se},~ O.~Ohlsson~Sax$^{d,}$\footnote{o.e.olssonsax@uu.nl}}\\
 \vskip  0.5cm

\small
{\em
$^{a}$ Institute of Cosmos Sciences and Estructura i Constituents de la Materia \\
Facultat de F\'isica, Universitat de Barcelona, Av. Diagonal 647, 08028 Barcelona, Spain
\vskip 0.05cm
$^{b}$Institut f\"ur Physik, Humboldt-Universit\"at zu Berlin, Newtonstra\ss e 15, D-12489 Berlin, Germany  
\vskip 0.05cm
$^{c}$ Department of Fundamental Physics, Chalmers University of Technology, 412 96 G\"oteborg, Sweden
\vskip 0.05cm
$^{d}$  Institute for Theoretical Physics and Spinoza Institute\\ Utrecht University, 3508 TD Utrecht, The Netherlands}
\normalsize
\end{center}

\begin{abstract}

We evaluate the exact one-loop partition function for fundamental strings whose world-surface ends on a cusp at the boundary of  $\ads_4$ and has a ``jump'' in $ \cp^3$.  This allows us to extract the stringy prediction for the ABJM generalized cusp anomalous dimension  $\Gamma_{\textrm{cusp}}^{\textrm{ABJM}}(\phi,\theta)$ up to NLO in sigma-model perturbation theory.
With a similar analysis, we present the exact partition functions for folded closed string solutions moving in the $\ads_3$  parts of $\ads_4\times \cp^3$ and $\ads_3\times \sphere^3\times \sphere^3 \times \sphere^1$ backgrounds. 
Results are obtained applying to the string solutions relevant  for the  $\ads_4$/CFT$_3$ and $\ads_3$/CFT$_2$ correspondence
the tools previously developed for their $\ads_5\times \sphere^5$ counterparts. 
\end{abstract}

\newpage
\tableofcontents

\section{Overview}

Since the seminal paper of~\cite{Gubser:2002tv} semiclassical quantization of strings  has been an efficient way to quantify the structure of the AdS/CFT correspondence~\cite{Maldacena:1997re,Witten:1998qj,Gubser:1998bc}, particularly in connection with the supposed underlying integrability of the planar system~\cite{Tseytlin:2010jv,McLoughlin:2010jw}.  In the case of the spectral problem for the $\ads_5$/CFT$_{4}$ correspondence~\cite{Beisert:2010jr}, a number of purely stringy-derived results have first inspired the formulation of the integrable equations for the spectrum and then confirmed the results obtained via an integrability-based approach.

The remarkable new findings of~\cite{Drukker:2012de, Correa:2012hh} establish, proving it via comparison with perturbative gauge theory results~\cite{Drukker:2011za,Correa:2012nk}, that suitably modified thermodynamic Bethe ansatz integral equations can efficiently calculate the generalized quark-antiquark potential $V(\lambda,\phi,\theta)$~\cite{Drukker:2011za}, or generalized cusp, realized as a family of Wilson loops with one geometric ($\phi$) and one ``internal'' ($\theta$) angle, which shows a non-trivial coupling dependence. 
Equivalently remarkable observations~\cite{Correa:2012at,Fiol:2012sg}  have further analyzed via localization the small angle region $\phi,\theta\ll 1$ of $V(\lambda,\phi,\theta)$ (related to a variety of other physical observables), which was then evaluated as an exact function of the coupling.    While confirming the intuition that the integrability of the system should apply also to the case of open strings and their dual Wilson loop configurations (see~\cite{Drukker:2005cu,Drukker:2006xg} for earlier studies in this sense),   these findings open the way to a wide variety of further developments.

In particular, we evaluate here the exact one-loop partition function of macroscopic Type IIA strings in $\ads_4\times \cp^3$ background with world-surface ending on a pair of (time-like) antiparallel lines respectively in flat space (section~\ref{sec:lines}) and in $\sphere^2\times \mathbb{R}$ (section~\ref{sec:cusp}), adding in the second case a ``jump'' in $\cp^3$. The study of the corresponding Wilson loops operators in three-dimensional $\mathcal{N}=6$ supersymmetric Chern-Simons theory (ABJM theory)~\cite{Aharony:2008ug,Benna:2008zy}, highly non-protected observables,  has  been only recently carried out in~\cite{Griguolo:2012iq}, while previous literature focused on supersymmetric~\cite{Drukker:2008zx,Berenstein:2008dc,Chen:2008bp,Rey:2008bh,Drukker:2009hy,Lee:2010hk,Cardinali:2012ru} or light-like~\cite{Henn:2010ps,Bianchi:2011rn,Wiegandt:2011uu} Wilson loops. The study of singularities, renormalization and exponentiation properties of these Wilson loops, contained in~\cite{Griguolo:2012iq}, has provided the ABJM extension (significantly more sophisticated) of the  $\mathcal{N}=4$ SYM ``quark-antiquark'' potential~\cite{Erickson:1999qv,Erickson:2000af}\cite{Pineda:2007kz} and,  respectively,  its generalized version~\cite{Drukker:2011za}.
Here we consider the corrections to their leading strong coupling behavior, with an  analysis similar to  the ones in~\cite{Forini:2010ek} and~\cite{Drukker:2011za}. The ``jump'' in $\cp^3$ can be realized on a generic $\sphere^1$ within $\cp^3$, since the R-symmetry of the theory makes the choice fully general.
Once the connection between localization and integrability is realized as in~\cite{Correa:2012at} also within the ABJM framework,\footnote{To this purpose, contours analog to the ones exploited in~\cite{Correa:2012at} (e.g. the solutions of~\cite{Drukker:2007qr} and others there cited) are in fact yet to be constructed.} we expect that our formula (\ref{cusp1loopexptheta0})  below will be reproduced via the strong coupling expansion of the exact results for the  $1/2$ BPS Wilson loop  found in~\cite{Drukker:2010nc}. As remarked in~\cite{Correa:2012hh}, the evaluation of the yet
undetermined function $h(\lambda)$ which appears in the ABJM magnon dispersion relation~\cite{Gaiotto:2008cg,Grignani:2008is,Nishioka:2008gz,Gromov:2008qe} could then finally become possible.\footnote{%
  For the leading order corrections to $h(\lambda)$ at strong coupling see~\cite{McLoughlin:2008ms,Alday:2008ut,Krishnan:2008zs,McLoughlin:2008he,Gromov:2008fy,Astolfi:2008ji,Bandres:2009kw, Abbott:2010yb,Abbott:2011xp,Astolfi:2011ju,Astolfi:2011bg}. At weak coupling the expansion of $h(\lambda)$ has been calculated to fourth order in the coupling constant~\cite{Minahan:2009aq,Minahan:2009wg,Leoni:2010tb}.
}%

It would be interesting to provide an analysis similar to the one here performed for the string configurations dual to the  1/6 BPS loops of~\cite{Drukker:2008zx, Rey:2008bh,Berenstein:2008dc}. The starting point would be there a classical string configuration represented by a solution ``smeared'' along a $\cp^1$.  As we know from the exact results of the ABJM matrix model~\cite{Drukker:2010nc}, the correspondingly more complicated ansatz will result in the same leading, classical, value as for the case here obtained but in a different, likely much more involved, fluctuation spectrum. In the recent~\cite{Kim:2012nd}, the partition function for the string solution corresponding to the ABJM circular Wilson loop has been evaluated, whose discrepancy with the exact prescription from the matrix model works similarly to the case of its $\ads_5\times \sphere^5$ counterpart~\cite{Kruczenski:2008zk}.

\bigskip
Below we also revisit the one-loop correction to the energy of  a string folded in the $\ads_3$ subspace of the $\ads_4\times \cp^3$ background, evaluating its exact partition function and deriving from it the relevant long and short string limits, as was done for the $\ads_5\times \sphere^5$ case in~\cite{Beccaria:2010ry}.
We recall that the genuinely field-theoretical tools used here for a direct world-sheet quantization are in principle alternative or complementary to the  integrability, algebraic curve approach to the semiclassical quantization of strings~\cite{Gromov:2007aq,Gromov:2008ec,Vicedo:2008jy, Gromov:2008bz}. This observation, in fact, would be highly non-trivial for the case of the Wilson minimal surfaces corresponding to the quark-antiquark potential~\cite{Maldacena:1998im} or its generalized version~\cite{Drukker:2011za} that we see here below in their ABJM extension. However, the recent results of~\cite{Janik:2012ws} establish at the classical level a well-defined description of such minimal surfaces in terms of appropriate algebraic curves,\footnote{The analysis of~\cite{Janik:2012ws} extends to the minimal surface corresponding to the Wilson loop with a null cusp and to the classical solutions corresponding to correlation functions of local operators. A possibly related contraction of minimal surfaces in $\ads$ starting from higher genus algebraic curves has been presented in~\cite{Ishizeki:2011bf}. }
thus giving confidence in a  possible extension to this case of the same powerful tools used to compute quantum corrections to  classical spinning string solutions~\cite{Gromov:2007aq,Gromov:2008ec,Vicedo:2008jy}, which might as well work in the ABJM setting.
While for closed spinning string solutions the algebraic curve approach is already an established powerful device for obtaining quantum corrections, a direct comparison of the results obtained is yet problematic, see section~\ref{sec:folded} below.
To clarify these discrepancies, as well as for a better general understanding of world-sheet quantization, it would be very interesting to develop the  necessary analytic  tools  for an exact analysis of the case in which the folded string is both spinning in $\ads_3$ and rotating in $\cp^3$. This highly non-trivial problem, however,  has not yet been resolved in the $\ads_5\times \sphere^5$ framework, as it would require the exact diagonalization of the mass matrices appearing in the study of  fluctuations.  As a starting point to this purpose, it might result useful our refined analysis of fluctuations around the folded string moving non-trivially in $\ads_3\times \sphere^1$~\cite{Frolov:2002av}, to which Appendix~\ref{app:ads5xs5} is devoted. There, we show how the known mixing of bosonic fluctuations occurring for non-vanishing  $\ads_3$ spin and  $\sphere^1$ orbital momentum~\cite{Frolov:2002av,Iwashita:2010tg}  has its supersymmetric counterpart in a non-trivial fermionic mass matrix.

We also report on similar results valid for the same configurations once they are seen as  Type IIB string solutions for the the sigma--model in a $\ads_3\times \sphere^3\times \sphere^3\times \sphere^1$ background with RR fluxes, and its limiting case  $\ads_3\times \sphere^3\times \torus^4$~\cite{Pesando:1998wm,Rahmfeld:1998zn,Park:1998un}.  
For this system a full set of Bethe equations for the spectrum were proposed~\cite{Babichenko:2009dk,OhlssonSax:2011ms}. The one-loop world-sheet analysis carried out in this paper can be seen as a first step in checking these equations beyond
leading order at strong coupling.

\bigskip

In the following two sections we report some exact one-loop results for the  relevant string solutions mentioned above in the two backgrounds of interest here, $\ads_4\times \cp^3$ and $\ads_3\times \sphere^3\times \sphere^3\times \sphere^1$. 
The details for their derivation, and several comments on the same details, are collected in the rich Appendices~\ref{app:AdS4xCP3} and~\ref{app:AdS3CFT2}.
Appendix~\ref{app:gamma_matrices} shortly describes the spinorial and Gamma matrices notation. In Appendix~\ref{app:ads5xs5} we revisit the analysis of fluctuations around the folded string  in $\ads_3\times \sphere^1$.

\section{Exact one-loop results in  $\ads_4 \times\cp^3$}
\label{sec:exactABJM}

In this section we evaluate the exact one-loop partition function of macroscopic Type IIA strings in $\ads_4\times \cp^3$ background with world-surface ending on a pair of (time-like) antiparallel lines respectively in flat space (section~\ref{sec:lines}) and in $\sphere^2\times \mathbb{R}$ (section~\ref{sec:cusp}), adding in the second case a ``jump'' in $\cp^3$.  
We continue in section~\ref{sec:folded} with the  evaluation of  the exact partition function for a folded string solution moving non trivially in the $\ads_3$ part of the background. Notation and relevant information for this background are presented in Appendix~\ref{app:AdS4xCP3}.
 
\subsection{Antiparallel lines}
\label{sec:lines}

In this case the relevant string solution found in the $\ads_5$ background~\cite{Maldacena:1998im} can be embedded within its $\ads_4$ subspace and the $\cp^3$ part of the sigma-model is classically irrelevant.\footnote{%
  In the dual gauge theory the scalar couplings are constant along the Wilson loop.%
} %
In the planar limit there is then no-difference with  the result obtained in~\cite{Maldacena:1998im}, provided the mapping $\lambda_{\ads_5}\to \bar\lambda= 2\pi^2\lambda$, with $\lambda=\frac{N}{k}$, is used.\footnote{This is general to all classical spinning string solutions of string theory in $\ads_5\times \sphere^5$, at least those with few charges~\cite{McLoughlin:2008he}, and to the known classical minimal surfaces relevant for Wilson loops~\cite{Drukker:2008zx}.} Therefore, the leading strong coupling value for the ``quark-antiquark'' potential in $\mathcal{N}=6$ Chern-Simons theory, as derived from the Nambu-Goto action for the dual sigma-model in $\ads_4\times \cp^3$ is\footnote{See~\cite{Conde:2011sw} for the leading order result  in a setup with flavors.} 
\begin{equation}\label{qqbarABJMlead}
  V_{\textrm{ABJM}}^{(0)}\,(\lambda,L) 
  = -\frac{\sqrt{\bar\lambda}\,\pi}{4\,\KK(\tfrac{1}{2})^2\,L}
  \equiv - \frac{\sqrt{\lambda}\,\pi^2}{2\sqrt{2}\,\KK(\tfrac{1}{2})^2\,L}, \qquad \lambda  \gg 1\,,
\end{equation}
where $L$ is the distance between the lines and $\KK$ is the complete elliptic integral of the first kind. 
The spectrum of bosonic quantum fluctuations over the classical solution follows very closely the formal analysis of~\cite{Drukker:2000ep}, the actual evaluation of~\cite{Chu:2009qt} and the analytical work elaborated in~\cite{Forini:2010ek}. Calculations are identical in nature  and result in a simple truncation of the bosonic spectrum  (one less transverse degree of freedom in the $\ads$ space) and in a slightly modified analysis for the  fermionic modes, which  follows from the more complicated nature of the compact space $\cp^3$.

Working in static gauge one finds two massive bosonic AdS fluctuations, one longitudinal and one transverse, together with six massless bosonic modes in $\cp^3$. After a global rotation and $\kappa$-symmetry gauge fixing reviewed in Appendix~\ref{app:lines}, the action for the two Type II A spinors becomes the standard one for two-dimensional Majorana fermions, and the resulting partition function is a rearrangement of the one in~\cite{Drukker:2000ep}
\begin{equation}\label{ZlinesABJM}
  \Gamma_{||}^{\textrm{ABJM}}= \frac{\det^{6/2}(-i\,\gamma^\alpha\,\nabla_\alpha+\sigma_3)}{\det^{1/2}(-\nabla^2+2)\,\det^{1/2}(-\nabla^2+R^{(2)}+4)\,\det^{4/2}(- \nabla^2)} \,,
\end{equation}
where $R^{(2)}$ is the scalar curvature, $\gamma^0=\sigma_2$, $\gamma^1=\sigma_1$ and $\gamma_0\gamma_1=-i\,\sigma_3$ are the Pauli matrices.
We can then immediately resort to the results of~\cite{Forini:2010ek}, writing down the effective action as a single integral in the  Fourier-transformed $\tau$-variable  ($\partial_\tau=i\,\omega$)
for a ratio of analytically exact functional determinants.\footnote{As explained in details in~\cite{Chu:2009qt, Forini:2010ek}, the fermionic operator can be further diagonalized after squaring it.} With the standard regularization, consisting in an explicit subtraction of divergences, the first correction in $\sigma$-model perturbation theory to the result (\ref{qqbarABJMlead}) reads 
\begin{equation}\label{Vlines}
V^{(1)}_{\textrm{ABJM}}(\lambda, L)=-\frac{1}{2}\,\frac{\pi}{\KK(\tfrac{1}{2})\,L}\lim_{\epsilon\to0}\int_{-\infty}^{+\infty}\frac{d\omega}{2\pi}\log\Big[\frac{\det^{6}\mathcal{O}_f ^\epsilon\,\omega^2\,\epsilon^2}{\det  \mathcal{O}_1\epsilon\,\det\mathcal{O}_2^\epsilon\,\det^{4}\mathcal{O}_0^\epsilon}\Big],
\end{equation}
in terms of the determinant expressions given in~\cite{Forini:2010ek}\footnote{See formulas (2.16)-(2.19) there.}. Explicitly, we have 
\begin{eqnarray}\label{VABJMan}
&&V^{(1)}_{\textrm{ABJM}}(\lambda,L)=-\frac{1}{2\,\KK(\tfrac{1}{2})\,L}\int_{0}^{+\infty} d\omega
\log\Big[\frac{2^{11} \omega ^{8} \sqrt{1-4 \omega ^4}\sqrt{1+\omega^4}}{\left(16 \omega ^4+1\right)^3}\Big]+\\\nonumber
&&~-\frac{1}{2\,\KK(\tfrac{1}{2})\,L}\int_{0}^{+\infty}\!\!\!\!\!\!\!\!\!\! d\omega
\log\left[\frac{ \sin
   ^6\Big[(1+i)\KK(\tfrac{1}{2})\, Z(\alpha_f)\,+\frac{\pi  \alpha_ f}{2 \KK(\tfrac{1}{2})}\Big]}{\sinh[2\,\KK(\tfrac{1}{2})\,Z(\a_1)]\,\sinh \Big[2(1+
   i)\KK(\tfrac{1}{2})\,Z(\alpha_2)+\frac{i \pi \alpha_2}{\KK(\tfrac{1}{2})}\Big]\,\sinh^{4}[2\,\KK(\tfrac{1}{2})\,\o]}\right]
\end{eqnarray}
where $Z$ is the Jacobi Zeta function and
\begin{equation}
  \alpha_1=\sn^{-1}\sqrt{1+2\,\o^2} \,, \qquad
  \alpha_2=\sn^{-1}\sqrt{1-i\,\o^2} \,, \qquad
  \alpha_f=\sn^{-1}\sqrt{1+4i\,\o^2} \,.
\end{equation}
The integral in the first line is easily done, while the second one is conveniently rewritten in terms of analytically known constants and a one-dimensional integral representation which can be numerically integrated with arbitrary high precision. The result is 
\begin{equation}\label{ABJMclassplusoneloop}
  V_{\textrm{ABJM}}(\sqrt{\lambda},L) 
  = -\frac{\sqrt{\lambda}\,\pi^2}{2\sqrt{2}\,\KK(\tfrac{1}{2})^2\,L} \, \Big[1+\frac{a_1}{\sqrt{\lambda}}+\mathcal{O}\Big(\frac{1}{(\sqrt{\lambda})^2}\Big)\Big] \,,
\end{equation}
with
\begin{equation}\label{a1number}
  a_1 = \frac{1}{3\sqrt{2}} - \frac{5\,\log2}{2\sqrt{2}\,\pi}
  +\frac{\sqrt{2}\,\KK(\tfrac{1}{2})}{\pi^2}\Big( 
  \frac{\KK(\tfrac{1}{2})}{2}-\frac{3\pi}{2\sqrt{2}}-\frac{\log 2}{\sqrt{2}}+\mathcal{I}^{\textrm{num}}
  \Big)=-0.3606061 \,,
\end{equation}
and $\mathcal{I}^{\textrm{num}}$ given in (\ref{numericsABJM}).

\subsection{Generalized cusp}
\label{sec:cusp}

Just as in the $\mathcal{N}=4$ SYM case~\cite{Drukker:2011za}, it turns out to be very useful to  add extra-parameters to the problem considered in the previous section~\footnote{For another interesting deformation of the system of antiparallel lines in $\mathcal{N}=4$ SYM see~\cite{Griguolo:2012gw}.}. The  antiparallel lines can be put on  $\sphere^2\times \mathbb{R}$   separated by an angle $ \pi-\phi$ along a big circle on $\sphere^2$. Using a conformal transformation, this is seen in flat $\mathbb{R}^{1,2}$ as two rays intersecting at a cusp, a geometrical angle $ \pi-\phi$ in the three-dimensional plane.\footnote{One can also see the cusp as the region where a straight line makes a sudden turn, the angle being then $\phi$.} One can also consider an internal angle, namely an angle in the SU(4) internal space. In the $\mathcal{N}=4$ SYM case, this is realized  when each line of the Wilson loop presents a different coupling to the scalars. A proposal for a dual string classical solution, given the well-defined interpretation for the coupling to the scalars  in terms of coordinates on $\sphere^5$,  naturally follows~\cite{Maldacena:1998im}
from the fact that the natural Wilson loop carries an SO(6) indices. There is not such an explicit prescription for the considerably more complicated coupling structure appearing in the  Wilson loops for $\mathcal{N}=6$ super Chern-Simons theory. Here, each of the half-BPS lines is defined as the as the holonomy of a \emph{superconnection}~\cite{Drukker:2009hy,Lee:2010hk}. Each ray breaks SU(4) to SU(3), selecting a direction in SU(4) in a non-trivial coupling to the bi-fundamental scalar fields~\footnote{The structure of the scalar coupling was understood already in~\cite{Drukker:2008zx, Berenstein:2008dc, Chen:2008bp, Rey:2008bh}.} and a coupling involving constant spinors~\cite{Drukker:2009hy,Lee:2010hk}.  We don't see how the  peculiar  and novel (with respect to $\mathcal{N}=4$ SYM) feature of a fermionic coupling might influence the analysis we perform here in terms of fluctuations over a bosonic classical solution. In total analogy with its $\ads_5\times \sphere^5$ counterpart, one can describe such finite discontinuity in the SU(4) internal space  as a ``jump'' in $\cp^3$,  which is natural to represent  in the form of a rigid (time-independent) ansatz for one of the six coordinates in $\cp^3$, with the other coordinates taking constant values.

One starts with global Lorentzian $\ads_4\times\cp^3$, the natural dual to gauge theory on $\sphere^2\times\mathbb{R}$,
\begin{equation}
  ds^2 = -\cosh^2\rho\,dt^2+d\rho^2+\sinh^2\rho\,d\varphi^2+ 4 ds^2_{\cp^3}
\end{equation}
where $ds^2_{\cp^3}$ is the metric (\ref{metricCP3}). 
Taking as world-sheet coordinates $t$ and $\varphi$, we choose (the translational invariance of the problem makes the embedding coordinates to be $\varphi$-dependent only)
\begin{equation}\label{ansatz}
  \rho=\rho(\varphi) \,, \qquad
  \theta_1=\theta_1(\varphi) \,, \qquad
  \alpha=\varphi_1=\theta_2=\varphi_2=\chi=0 \,,
\end{equation}
where the AdS coordinate $\varphi$ varies in the domain $[\phi/2,\pi-\phi/2]$ and the $\C P$ coordinate $\theta_1$ in $[-\theta/2,\theta/2]$. 
Notice that the choice of the ansatz for the $\cp^3$ coordinates is fully general. The ways to select an arc in $\cp^3$ are all equivalent, in that there is always a global SU(4) rotation that maps them to the same form in terms of the embedding coordinates. In particular, in analyzing fluctuations, this allows one to work with different coordinates in $\cp^3$ which are more suitable for calculations.

\subsubsection{The classical limit}

The ansatz (\ref{ansatz}) solves the classical equations of motion the same way as in the $\ads_5\times \sphere^5$ case. This holds  in the case of a non-trivial ansatz for any of the $\cp^3$ coordinates.%
\footnote{This happens provided the ansatz is suitably chosen. In the case of, say,  $\psi_i, i=1,2,3$ in (\ref{metricCP3}), a rescaled classical ansatz will compensate the factor 4 in front of the total metric (\ref{total_metric}).} To account for this generality we call generically  $\vartheta\equiv\vartheta(\varphi)$ the non-trivial  coordinate describing the jump.
Thus, we can exploit the classical results of~\cite{Drukker:2011za}\footnote{Without even the need to change the variable names!}  with the usual mapping $\lambda\to\bar\lambda$.  
In particular, from (\ref{classical-action}) one finds, at leading order in the $\bar\lambda\to\infty$ limit 
\begin{equation}\label{classical-action-2}
  {\Gamma_{\textrm{cusp}}^{\textrm{ABJM}}}^{(0)}(\phi,\theta) \equiv\frac{\mathcal{S}^{\textrm{reg}}_\text{cl}}{T}= \frac{\sqrt{\bar{\lambda}}}{2\pi} \frac{2\sqrt{b^4+p^2}}{b\,p}
  \left[\frac{(b^2+1)p^2}{b^4+p^2} \KK(k^2)-\EE(k^2)\right] \,,
  \qquad \bar\lambda\gg 1 \,.
\end{equation}
In the limit $\phi\to\pi$ and $\theta=0$ (for which is $k^2=1/2$)  this  gives
\begin{equation}
  \lim_{\phi\to\pi}{\Gamma_{\textrm{cusp}}^{\textrm{ABJM}}}^{(0)}(\phi,\theta)
  = -\frac{\sqrt{\bar\lambda}\,\pi}{4\,\KK(\tfrac{1}{2})^2\,(\pi-\phi)}\label{ads-coinc} \,,
\end{equation}
which obviously agrees with  (\ref{qqbarABJMlead})  with the replacement $\pi-\phi\to L$.


Since the classical Lagrangian is the same, the BPS condition is still  $E=\pm J$~\cite{Drukker:2007qr}\footnote{See the discussion on the BPS bound in Appendix~C.2  of~\cite{Drukker:2007qr}.} and implies  $\phi=\pm\theta$.
In particular, we can expand around the straight-line configuration $\theta=\phi=0$, exploiting (\ref{phithetapinfinity}) and the results of~\cite{Drukker:2011za} to get~\footnote{We thank M. Bianchi, L. Griguolo and D. Seminara for pointing out that a factor 1/2 was missing in \eqref{V0ads-expand} in the previous version of this paper. See~\cite{toappear}, to appear.}
\begin{eqnarray}
\nonumber
\!\!\!\!\!\!\!&&{\Gamma_{\textrm{cusp}}^{\textrm{ABJM}}}^{(0)}(\phi,\theta)
=\,\frac{\sqrt{\bar{\lambda}}}{2\pi}\,\Big[
\frac{1}{2\pi}(\theta^2-\phi^2)
-\frac{1}{16\pi^3}(\theta^2-\phi^2)\left(\theta^2-5\phi^2\right)
+\frac{1}{128\pi^5}(\theta^2-\phi^2)\left(\theta^4-14\theta
^2\phi^2+37\phi^4\right)\\\label{V0ads-expand}
&&\qquad\qquad\qquad\qquad\qquad-\frac{1}{40968\pi^7}(\theta^2-\phi^2)
\left(\theta^6-27\theta^4\phi^2+291\theta^2\phi^4-585\phi^6\right)
+\order((\phi,\theta)^{10})\Big]\,.
\end{eqnarray}


\subsubsection{Fluctuations and one-loop correction to $\Gamma_{\textrm{cusp}}^{\textrm{ABJM}}$}

The analysis of fluctuations, reviewed in Appendix~\ref{app:cusp}, goes much the same way as in the $\ads_5\times \sphere^5$ case~\cite{Drukker:2011za}, presenting additional features observed for the ABJM folded string  (see~\cite{McLoughlin:2008he} and section~\ref{sec:folded} below), as for example the splitting of bosonic fluctuations in $\cp^3$ into ``heavy'' and ``light'' degrees of freedom.
As in~\cite{Drukker:2011za}, we present here the relevant solutions in the two cases in which both bosonic and fermionic fluctuations decouple, and the partition function can be evaluated in an exact way.
In the case of $\theta=0$, one gets a simple redistribution of the  determinants appearing in the $\ads_5\times \sphere^5$ case, see formula (\ref{reg_effectiveaction_theta0}). 
In the case of $\phi=0$, the slightly more complicated structure of $\cp^3$ plays its role, and one finds a splitting in light bosons and light fermions resulting in the partition function (\ref{reg_effectiveaction_phi0}). We will evaluate below the one-loop correction to the value of the generalized cusp in (\ref{classical-action-2}) defined as
\begin{equation}
{\Gamma_{\textrm{cusp}}^{\textrm{ABJM}}}^{(1)}(\phi,0)=\frac{\Gamma^{\textrm{reg}}_{\theta=0}}{T}\,,\qquad
{\Gamma_{\textrm{cusp}}^{\textrm{ABJM}}}^{(1)}(0,\theta)=\frac{\widetilde{\Gamma}^{\textrm{reg}}_{\phi=0}}{T}\,.
\end{equation}
In each of the cases above one derives a one-dimensional integral over an expression involving complicated special elliptic 
functions. As a first relevant check of the first expression above, the light-like cusp limit ${\Gamma_{\textrm{cusp}}^{\textrm{ABJM}}}^{(1)}(\phi=i\,\infty, 0)$ matches~\cite{inprogress} the strong coupling value of the so-called scaling function, or cusp anomalous dimension, as found in~\cite{McLoughlin:2008ms, Alday:2008ut, Krishnan:2008zs}, see formula \eqref{E1largespin} below.
 The results for the antiparallel lines, (\ref{ABJMclassplusoneloop})-(\ref{a1number})  are reproduced from the partition function for $\theta=0$ (\ref{reg_effectiveaction_theta0}) in the limit $\phi\to\pi$, having set $k^2=1/2$ and having taken into account the different normalizations for the partition functions~\footnote{One can check that this means to subtract to the numerical result of (\ref{reg_effectiveaction_theta0}) the constant $\KK(\textstyle{\frac{1}{2}})/{2\sqrt{2}\pi}$, in other words the value for the constant $a_1$ in (\ref{ABJMclassplusoneloop}) obtained with the prescription here used and described in Appendix \ref{app:theta0} below is
\be\label{a1numberbis}
a_1|_{\rm reg}=\frac{1}{3\sqrt{2}} - \frac{5\,\log2}{2\sqrt{2}\,\pi}
  +\frac{\sqrt{2}\,\KK(\tfrac{1}{2})}{\pi^2}\Big( 
  \frac{\KK(\tfrac{1}{2})}{2}+\frac{\pi}{4}(1-3\sqrt{2})-\frac{\log 2}{\sqrt{2}}+\mathcal{I}^{\textrm{num}}
  \Big)=- 0.15194941\,,
\ee 
which differ from the result (\ref{a1number}) for the constant mentioned above.}.
This matching works exactly as in the $\ads_5\times \sphere^5$ case~\cite{Drukker:2011za} with the reproduction of the results in~\cite{Forini:2010ek}. In particular,  
the ratio between the world--sheet time $\cal T$ and  the target space time $T$, which  furnishes the expected $1/(\pi-\phi)$ pole for generic $k$
\begin{equation}
\frac{\mathcal{T}}{T}=\frac{1}{\sqrt{k}(1-k^2)^{1/4}\sqrt{p}}
=\frac{2}{\pi-\phi}\frac{\EE-(1-k^2)\KK}{k\sqrt{1-k^2}}\,.
\end{equation}
becomes, in the  $k^2=1/2$ case,  $\pi/(\pi-\phi)\KK(1/2)$. Replacing $\pi-\phi\to L$, this is indeed the 
rescaling used in~\cite{Chu:2009qt,Forini:2010ek} for the classical $\ads_5\times \sphere^5$ solution which has been exploited in the derivation of (\ref{ABJMclassplusoneloop}).  

Particularly interesting are the small $\phi$ (for $\theta=0$) and small $\theta$ (for $\phi=0$) limits, translated in expansions at small elliptic modulus. In these cases the integrand becomes a series of hyperbolic functions which can be exactly integrated resulting in explicit analytic results, as derived in the expansions (\ref{expansiontheta0}) and (\ref{expansionphi0}) using (\ref{phithetapinfinity}). The result for the $\theta=0$ case reads
 \begin{equation}
\label{cusp1loopexptheta0}
 {\Gamma_{\textrm{cusp}}^{\textrm{ABJM}}}^{(1)}(\phi,0)
=
\frac{7}{8}\frac{ \phi^2}{ 4\pi^2} +\Big(\frac{127}{32}-\frac{9 \zeta (3)}{4}\Big)\frac{\phi^4}{16\,\pi^4} +\Big(\frac{527}{32}-\frac{39 \zeta (3)}{8}-\frac{45 \zeta (5)}{8}\Big) \frac{\phi^6}{64\,\pi^6} + \order(\phi ^8)\,,
\end{equation} 
Notice that, as in the $\ads_5\times \sphere^5$ case, a pattern of uniformly increasing transcendentality appears.
The first terms in the expansion above should be reproduced via the strong coupling expansion of the exact results for the  $1/2$ BPS Wilson loop  found in~\cite{Drukker:2010nc}, once the relevant connection~\cite{Correa:2012at} with the small angle region $\phi,\theta\ll 1$ of $\Gamma_{\textrm{cusp}}$ is generalized to the ABJM case.
The same coefficients should also be reproduced by a strong coupling analysis of the ABJM-generalized BTBA equations of~\cite{Drukker:2012de,Correa:2012hh}, once they become available.

We also present here our attempt of evaluation for the small $\theta$ expansion in the $\phi=0$ case, based on the analysis carried on in Appendix \ref{app:phi=0}
\begin{eqnarray}\label{cusp1loopexpphi0}
{\Gamma_{\textrm{cusp}}^{\textrm{ABJM}}}^{(1)}(0,\theta)
&&= -\frac{3}{8}\frac{\theta^2}{ 4\pi ^2} - \Big(\frac{1}{32}+\frac{\log (2)}{2}+\frac{3 \zeta (3)}{4}\Big) \frac{\theta^4}{16\,\pi^4}+ \\\nonumber
 &&\qquad\qquad\qquad\qquad\qquad+ \left(\frac{7}{32}+\frac{\log (2)}{2}-\frac{15 \zeta (5)}{8}\right) \frac{\theta^6}{64\,\pi^6} +\order(\theta^{8})\,.
\end{eqnarray}
This result is in disagreement with the observation of~\cite{Griguolo:2012iq}~\footnote{ The results of~\cite{Griguolo:2012iq} were not yet available when the earliest preprint of this paper was made public. See section 3.2 there.} that the BPS condition for the ABJM generalized cusp stays the same as in the $\mathcal{N}=4$ SYM case, being  $\theta=\pm \phi$ and implying therefore (under the assumption of analyticity for $\Gamma_{\textrm{cusp}}(\phi, \theta)$) that the coefficient of $\phi^2$  in (\ref{cusp1loopexptheta0}) should be the same as the coefficient of $\theta^2$ in (\ref{cusp1loopexpphi0}). It is not clear to us how to obtain an expansion compatible with the study of~\cite{Griguolo:2012iq}, and it is likely that an alternative or more careful analysis will lead to correct  (\ref{cusp1loopexpphi0})~\footnote{This would also likely eliminate the odd presence of $\log 2$ terms in it.}. In particular, it would be interesting to investigate the possible dependence of this discrepancy on the choice of regularization. While the regularization scheme used in this paper seems natural from the world-sheet point of view, the distinction between ``light'' and ``heavy'' modes modes in the spectrum (a feature of any non trivial ``motion'' in $\cp^3$) might suggest a prescription that treats the UV divergencies of the corresponding determinants differently, much in the same way as in the discussion of~\cite{Alday:2008ut, Abbott:2011xp}. It is likely that the use of algebraic curve methods (when developed at this order of sigma-model perturbation theory, see Overview) would help in clarifying possible regularization issues~\footnote{See the discussion around C.42 below for the case of the spinning string in $\ads_3$.}  (for example on the lines of~\cite{Abbott:2010yb} and~\cite{Abbott:2012dd}).

\subsection{Folded  string}
\label{sec:folded}

In this section we report what the semiclassically ``exact'' analysis used in the previous section can teach us if we use it for the quantization of the folded string in $\ads_4 \times\cp^3$~\cite{Krishnan:2008zs, Gromov:2008fy,McLoughlin:2008ms,McLoughlin:2008he}, recently reconsidered in~\cite{LevkovichMaslyuk:2011ty,Beccaria:2012qd}\footnote{See also the recent~\cite{LopezArcos:2012gb} for an interesting interpretation of the results in~\cite{McLoughlin:2008he}.}. In particular, we consider the case when the string is spinning in the $\ads_3$ part of $\ads_4 \times\cp^3$. We then have\footnote{We follow the notation of~\cite{McLoughlin:2008he} and set the $\ads_4$ radius to 1.} 
\begin{eqnarray}
  \nonumber
  &&ds^2_{\ads_3}= -\cosh^2\rho dt^2+d\rho^2 +\sinh^2\rho\, d\phi^2\\\label{solfolded}
  &&t=\kappa\tau \,, \quad \phi=\omega \tau\,, \quad \rho(\sigma)=\rho(\sigma+2\pi) \,,\\\nonumber
  &&
  \psi_1= {\pi\over 4}\,, \quad
  \psi_2={\pi\over 2}\,, \quad
  \psi_3={\pi\over 2}\,, \quad
  \tau_s=0 \,, \,\,s=1,2,3.
\end{eqnarray}
In this case, and in  static gauge, both bosonic and fermionic fluctuations are decoupled and an analytically exact one-loop partition function can be written down. Since the classical motion of the string is restricted to  $\ads_3$, in the planar limit and classically  the result is the same as the one obtained in~\cite{Gubser:2002tv,Frolov:2002av}, the only requirement being the mapping $\lambda_{\ads_5}\to \bar\lambda= 2\pi^2\lambda$ is used.\footnote{As observed in~\cite{McLoughlin:2008he}, the equivalence of the classical solutions for the two backgrounds $\ads_5\times \sphere^5$ and $\ads_4 \times \cp^3$ extends to the results for the classical energy and spin. Therefore at the classical level the prescription relating the $\ads_5\times \sphere^5$ and $\ads_4 \times \cp^3$ results is equivalent  to no rescaling at all. Namely, one rescales  $\lambda_{\ads_5}\to 2 \bar\lambda$,  and  replaces $E, S$ in $\ads_5$ results by $2E$ and $2S$. Changing the string tension by $2$ is compensated by rescaling of charges by $2$ so that classical parameters remain the same.}

The semiclassical analysis of fluctuations is again only a slight modification of the one conducted in the $\ads_5\times \sphere^5$ framework~\cite{Beccaria:2010ry},  and in fact an analog set of information can be inferred in the two relevant limits of long and short strings via a systematical expansion in large and small spin.
For the bosonic spectrum we work in static gauge, since it is in this case that they are all decoupled  and an analytic expression for the one-loop partition function can be derived. The result is obtained by truncating the fluctuations in $\ads_5$~\cite{Frolov:2002av} so that one transverse degree of freedom is eliminated. There are then additional  six massless degrees of freedom in $\cp^3$.
As reviewed in Appendix~\ref{app:foldedABJM}, the bosonic contribution to the one-loop effective action is then
\begin{gather}\label{gammabosonsfolded}
  \widetilde{\Gamma}_B=\frac{1}{2}\,\Big(\log\det\mathcal{O}_{\bar\phi} + \log\det\mathcal{O}_\theta+6\,\log\det\mathcal{O}_0\,\Big) \,,
  \\\label{bosonoperatorsABJM}
  \mathcal{O}_{\bar \phi} = -\d_\s^2+\Omega^2+  2 \r'^2  + \frac{ 2 \k^2 \o^2 }{ \r'^2} \,, \qquad
  \mathcal{O}_\theta = -\d_\s^2+\Omega^2+2\r'^2 \,, \qquad 
  \mathcal{O}_0 =  -\d_\s^2+\Omega^2 .  
\end{gather}
From the analysis of Appendix~\ref{app:foldedABJM}, it follows that the fermionic counterpart to (\ref{gammabosonsfolded}) in the effective action reads 
\begin{gather} \label{fer}
 \widetilde\Gamma_F = -\frac{1}{2} \Big( 3 \log \det \mathcal{O}_{\psi_+} + 3 \log \det \mathcal{O}_{\psi_-}+2\,\mathcal{O}_0 \Big) \,, 
 \\\label{fermionicoperatorsABJM}
 \mathcal{O}_{\psi_\pm} = -\partial_\s^2+\Omega^2  +\mu^2_{\psi_\pm} \,,\qquad
 \mu^2_{\psi_\pm}= \pm \r'' + \r'^2 \,, \qquad
 \mathcal{O}_0=-\partial^2_\sigma+\Omega^2 \,.
\end{gather}
Of the eight effective fermionic degrees of freedom, two are massless and the other ones have masses which coincide with the ones appearing, although with different multiplicities,  in the $\ads_5\times \sphere^5$ case~\cite{Beccaria:2010ry}. The same holds for the bosonic fluctuations (\ref{bosonoperatorsABJM}).
It is therefore easy to exploit the analysis of~\cite{Beccaria:2010ry} and write down immediately the  one-loop correction to the 
energy of the folded spinning string in the present case\footnote{Using the large-$\Omega$ expansions of the determinants, also written down in~\cite{Beccaria:2010ry}, it is easy to check that the expression (\ref{final}) is UV finite.}
\begin{equation}\label{final}
  E_1=-\frac{1}{4\pi\kappa} \int_{-\infty}^{\infty} \mathrm{d}\Omega\, 
  \log\frac{\det^6 \mathcal{O}_\psi}{\det\mathcal{O}_{\phi}  \, \det\mathcal{O}_{\theta} \, \det^4\mathcal{O}_0} \,,
\end{equation}
where $\k=\frac{2k }{\pi}\,\KK$, $\KK= \KK(k^2)$ (see (\ref{kappaomega})) and we
used that $\det\mathcal{O}_{\psi_+} = \det
\mathcal{O}_{\psi_-}\equiv\det\mathcal{O}_\psi$~\cite{Drukker:2011za}. Above, the
determinants have the following explicit expressions\footnote{\label{antiperiodicbc}
  The fermionic functional determinants are derived, as in~\cite{Beccaria:2010ry},
  choosing \emph{periodic} boundary conditions. We agree therefore with the
  discussion in~\cite{Beccaria:2010zn}, according to which for the folded string
  (in $\ads_5\times\sphere^5$ and in $\ads_4\times\cp^3$) it is not clear
  whether antiperiodic boundary conditions should be used for the
  fermions~\cite{Mikhaylov:2010ib}. We notice that this would change the
  $\cosh^2$ in $\sinh^2$ in the fermionic functional determinant
  (\ref{detpsi2}), cf. footnotes \ref{footnote1} and \ref{footnote2}. %
} %
\begin{align}\label{detbeta2}
  \det\O_\theta &= \phantom{-} 4\sinh^2 \Big[2\KK Z(\alpha_\theta\,|\, k^2) \Big]
  &\text{where}&&
  \sn(\alpha_\theta\,|\,k^2) &= \frac{\sqrt{1 + k^2 + \big(\frac{\pi\,\Omega}{2\KK}\big)^2}}{k} \,,
  \\\label{detphi2}
  \det\O_\phi &= \phantom{-} 4\sinh^2\Big[2\widetilde{\KK} \, Z(\alpha_\phi\,|\, \tilde{k}^2)\Big]
  &\text{where}&&
  \sn(\alpha_\phi\,|\, \tilde{k}^2) &=\frac{\sqrt{1 + \big(\frac{\pi\,\Omega}{2\widetilde{\KK}}\big)^2}}{\tilde{k}} \,,
  \\\label{detpsi2}
  \det\O_\psi &=-4\cosh^2\!\Big[\widetilde{\KK} \, Z(\alpha_\psi\,|\, \tilde{k}^2)\Big]
  &\text{where}&&
  \sn(\alpha_\psi\,|\, \td{k}^2) &= \frac{\sqrt{1 + \big(\frac{\pi\, \Omega}{\widetilde{\KK}}\big)^2}}{\tilde{k}} \,,
  \\\label{dettrivial}
  \det\O_0 &= \phantom{-} 4\sinh^2\Big[\pi\Omega\Big] \,,
\end{align}
where $\tilde{k}^2=\frac{4k}{(1+k)^2}$ and $\widetilde{\KK}=\KK\big(\tilde{k}^2\big)$.
 
\subsubsection{Long and short string limits}
 \label{sec:long-and-short-ads4}
 
The expression  (\ref{final}), as such only  numerically evaluable for given values of $k$, gives valuable analytic information once the integrand is expanded in the  large spin (``long string'' or $k\to 1$) limit 
or in the small spin (``short string''  or $k\to 0$) limit.  
In the \emph{long string limit}, the final result for the large spin expansion, obtained following the lines of~\cite{Beccaria:2010ry}\footnote{In particular, lacking a  complete control over them, exponential corrections occurring in the large spin expansion of the determinants are systematically neglected, see Appendix~E in~\cite{Beccaria:2010ry}.}, reads \begin{equation}\label{E1large}
  E_1 = \frac{\kappa_0}{\kappa} \,
  \bigg[\Big(c_{01}\,\kappa_0 + c_{00} + \frac{c_{_{0,-1}}}{\kappa_0}\Big) + \Big(c_{11}\,\k_0+c_{10} + \frac{c_{_{1,-1}}}{\kappa_0}\Big) \eta+\order(\eta^2) \bigg] \,,
 \end{equation}
where the explicit values are 
\begin{align}\label{coefficientsfirst}
  c_{01} &= -5\frac{\log 2}{2} \,, &
  c_{00} &= \frac{1}{2} +\frac{5}{\pi} \log 2 \,, &
  c_{_{0,-1}} &= -\frac{1}{3}\,, 
  \\\label{coefficientslast}
  c_{11} &= 0 \,, &
  c_{10} &= -\frac{5}{2\pi} \log 2 \,, &
  c_{_{1,-1}} &= \frac{1}{4\pi }+\frac{5\log2}{2\pi^2}\,.
\end{align}
For completeness, we report here the first few orders in the large spin expansion of the one-loop energy as found explicitly in terms of the spin (\ref{ES})%
\footnote{\label{footnote1}The ${1\over \log\S}$ term in \eqref{E1largespin} is consistent with the low-energy effective theory for an $\ads_4\times \C P^3$ folded string with periodic boundary conditions for fermions. Indeed, in the low-energy limit the energy is $E_{\rm le}= {c \pi\over 12 \log\S}$, with $c$ the central charge, which, in this case, is $c=6-2=4$, counting 6 massless bosons and 1 massless Dirac fermion with periodic boundary conditions.  B. Basso recently suggested, from considerations based on the Bethe ansatz, that massless fermions should have anti-periodic boundary conditions. We explicitly checked that changing  {\it only} the massless fermions from periodic to anti-periodic leads to a term ${7 \pi \over 12 \log\S}$ in agreement  with $E_{\rm le}= {c_{\rm AP}\, \pi\over 12 \log\S}$, where $c_{\rm AP}= 6+ 1=7$. However, such a choice seems unnatural from the point of view of a world-sheet analysis. We thank B. Basso for discussions on this point.}
\begin{align}\label{E1largespin}
  E_1 = -\frac{5\log2}{2\pi} \log\bar{\S} & + \frac{\pi + 10\log2}{2\pi} - \frac{\pi}{3\log\bar{\S}} 
  \\\nonumber
  & - \frac{1}{\bar\S} \Big[ \frac{20\log2}{\pi} \log\bar{\S} - \frac{2\pi + 30\log2}{\pi} + \frac{4\pi}{3\log^2\bar\S} \Big]
  +\O\bigg(\frac{1}{\bar{\S}^2}\bigg) \,,
   \qquad \S \gg 1 \,,
\end{align}
where $\bar{\S}=8\pi\S$.
 
With the expressions  (\ref{E1large})-(\ref{coefficientslast}) at hand, we are able to check the pattern of \emph{reciprocity} relations between terms subleading in the large $\cal S$ expansion which has been largely observed in the $\ads_5\times \sphere^5$ framework (for a review see~\cite{Beccaria:2010tb}), and that at the semiclassical string level has been always performed on an action of the type (\ref{E1large}). 
At the classical level this is already proven~\cite{Beccaria:2008tg}. To see that explicitly~\cite{Beccaria:2008tg,Beccaria:2010ry} at one loop we check the parity  under $\mathcal{S}\to -\mathcal{S}$ of the function $\mathcal{P}(\mathcal{S})$ defined perturbatively via  
\begin{equation}\label{defP}
\Delta(\mathcal{S})  =  \mathcal{P}(\mathcal{S} + \frac{1}{2}\,\Delta(\mathcal{S})), \qquad
\Delta(\mathcal{S}) = \Delta_0(\mathcal{S})  + \frac{1}{\sqrt\lambda}\,\Delta_1(\mathcal{S}) + \cdots~.
\end{equation}
where $\Delta_0(\mathcal{S}) = \mathcal{E}_0-\mathcal{S}$, and  $\Delta_1(\mathcal{S}) = \mathcal{E}_1$ are redefinitions of the classical and one-loop energies ($E_1=\sqrt\lambda \,\mathcal{E}_1$).  Solving perturbatively for $\mathcal{P}$ in (\ref{defP})
\begin{equation}
\mathcal{P}(\mathcal{S}) = \sum_{k=1}^\infty\frac{1}{k!}\left(-\frac{1}{2}\frac{d}{d\mathcal{S}}\right)^{k-1}[\Delta(\mathcal{S})]^k, \qquad
\mathcal{P} = \mathcal{P}_0 + \frac{1}{\sqrt\lambda}\,\mathcal{P}_1 + \cdots,
\end{equation}
one finds a parity invariant expression
\begin{equation}\label{P1ABJM}
  \mathcal{P}_1=-\frac{5 \log (2) \log\bar{\S}}{2 \pi }-\frac{\pi }{3 \log\bar{\S}}+\frac{1}{2}+\frac{5 \log (2)}{\pi}+\order(\bar{\S}^{-2})\,.
\end{equation}
The analysis above is a strong coupling confirmation of the  weak coupling analysis of~\cite{Beccaria:2009ny}, in which reciprocity relations were checked, up to four loops, for the large spin expansion of the anomalous dimensions of operators of  twist $L=1$  in the three-dimensional $\mathcal{N}=6$  super Chern-Simons theory.

Similarly, one can consider  the small spin or \emph{short string limit}~\cite{Tirziu:2008fk} of the folded string energy,  realized by  sending $\eta\to\infty$ or $k\to 0$.
To this purpose it is useful to consider an alternative expression for the final effective action,  which is free from IR-divergence problems in the relevant expansion, and that is reported in (\ref{final2}). Using the results there mentioned, one can write down the following small spin expansion of the one-loop correction
to the energy 
\begin{gather}\label{smallspin} 
  E_1 = E_1^{(\textrm{an})} + E_1^{(\textrm{nan})} \,, \qquad E_1^{(\textrm{nan})}=\frac{1}{2}+\order(\S) \,, \\
  E_1^{(\textrm{an})}=\sqrt{2\,\mathcal{S}}\,\bigg(1-3\,\log2+\frac{3}{16}\,\S\,\big(-5+6\log2+3\,\zeta(3)\big)+\order(\S^2)\bigg)\,,\\
  E=  \sqrt{2 S \sqrt{\lambda}} \bigg( 1+ \frac{\frac{3}{8}S + 1 - 3\log2}{\sqrt{\lambda}} + \dotsb \bigg) + \frac{1}{2}+\dotsb \,.
\end{gather}
As in~\cite{Beccaria:2010ry}, we have separated $E_1$  
into  an ``analytic'' part (with $\S$-dependence  similar to the 
 classical energy (\ref{Ec})) and  a ``non-analytic'' part, containing 
 would-be IR-singular  contributions of the   lowest eigenvalues. In the last
 line we have made explicit the $\lambda$-dependence, adding together the classical and one-loop results.%
 \footnote{\label{footnote2}The choice of anti-periodic boundary conditions for the massless fermions breaks the small spin expansions giving rise to divergent ${1\over \sqrt{\S}}$ terms.
}
The pattern of transcendental coefficients $\log 2,\zeta(3), \zeta(5)$  in the formula above resembles the one appearing in the $\ads_5\times \sphere^5$ case~\cite{Beccaria:2010ry, Beccaria:2010zn}. It would be nice to analyze the short string limit in a larger class of string configurations to see whether universality patterns similar to the ones observed for the $\ads_5\times \sphere^5$ case in~\cite{Beccaria:2012xm} manifest themselves also in this context. 
Our results are obtained using \emph{periodic} boundary conditions for fermions (see footnote \ref{antiperiodicbc}), the alternative choice of antiperiodic boundary conditions would, as in the $\ads_5\times \sphere^5$ case~\cite{Beccaria:2010zn,Beccaria:2012xm}, remove the $\log 2$ terms.
  
 As mentioned  in the Overview, relating the results obtained above with the one exploiting the algebraic curve tools is in general problematic. In particular, it is  not possible to compare our (\ref{E1largespin}) and (\ref{smallspin}) with the limit $\mathcal{J}\to0$ ($\ell\to0$) of the recent interesting results of~\cite{Beccaria:2012qd}. The ``non-analyticity'' of this limit has been first noticed in~\cite{Roiban:2009aa} and confirmed in~\cite{Roiban:2011fe,Gromov:2011de} (see also related discussion in~\cite{Fioravanti:2011xw}). A direct quantization of the folded string solution keeping  exact both the spin and the orbital momentum $(S,J)$, which would help elucidating this and similar problems of direct comparison, is a technically challenging problem that would be very interesting to solve.


\section{Exact one-loop results for folded strings in $\ads_3 \times \sphere^3 \times \sphere^3\times \sphere^1$}
\label{sec:intro_ads3_cft2}

In this section we will consider Type IIB superstrings on $\ads_3 \times
\sphere^3 \times \mbox{M}_4$, where the second compact manifold $M_4$ is either
$\sphere^3 \times \sphere^1$ or $\torus^4$.\footnote{%
  The case $M=K3$, which can be treated as an orbifold of $\ads_3\times
  \sphere^3\times \torus^4$, results in a straightforward generalization of the
  $\torus^4$ case.%
} %
The backgrounds we consider preserve 16 supersymmetries and are dual to
two-dimensional conformal field theories with $\mathcal{N}=(4,4)$ superconformal
symmetry. There are two inequivalent such superconformal algebras, referred to
as \emph{small} and \emph{large} $(4,4)$ superalgebras, with the maximal
finite-dimensional sub-algebras being $\algPSU(1,1|2)^2$ and
$\alg{d}(2,1;\alpha)^2$, respectively.  The algebras $\algPSU(1,1|2)^2$ and
$\alg{d}(2,1;\alpha)^2$ make up the super-isometries of the $\ads_3 \times
\sphere^3 \times \torus^4$ and $\ads_3 \times \sphere^3 \times \sphere^3 \times
\sphere^1$ backgrounds. In the latter the parameter $\alpha$, taking values $0 <
\alpha < 1$, is related to the radii of the two three-spheres
(see~\eqref{alpha}). In particular in the limits $\alpha \to 0$ and $\alpha \to
1$ the radius of one of the two three-spheres diverges. The corresponding
directions then become flat. Together with the $\sphere^1$ they can then be
compactified to a $\torus^4$. Hence, the case $M_4 = \torus^4$ can be obtained
as a limit of the more general case $M_4 = \sphere^3 \times \sphere^1$. For
$\alpha=1/2$ the radius of the two spheres are equal. The isometry is then given
by $\alg{d}(2,1;1/2)^2 = \algOSp(4|2)^2$.

Classical string theory in the above backgrounds can be formulated in terms of
integrable super-coset
sigma-models~\cite{Babichenko:2009dk}. In~\cite{Babichenko:2009dk,OhlssonSax:2011ms} Bethe
equations were proposed describing the string theory spectra, or equivalently
the spectrum of local operators in the dual CFT:s. Little is known about the CFT
dual of the string theories considered here.\footnote{%
 For discussions on the $\mathcal{N}=(4,4)$ $\cft_2$ duals see~\cite{Elitzur:1998mm,Gauntlett:1998kc,Cowdall:1998bu,Boonstra:1998yu,deBoer:1999rh,Papadopoulos:1999tw,Giveon:2003ku,Gukov:2004ym}.%
} %
However, if the proposed Bethe equations are correct they may provide some
insight to the spectra of the theories. The one-loop calculations performed in
this paper can be seen as a first step in checking these equations beyond
leading order at strong coupling.

In this section we evaluate the exact one-loop partition functions of a folded
string carrying angular momentum in $\ads_3$. This solution and its
generalization to a spinning string with non-zero angular momentum also on the
two three-spheres is written down in Appendix~\ref{app:AdS3CFT2}, which also
contains more details about the fluctuation spectrum. The appendix additionally
contains a one-loop analysis of the large spin scaling limit of the spinning
string, as well as a discussion of the embedding of the classical string
solution discussed in section~\ref{sec:cusp} into $\ads_3$.


\subsection{Fluctuations}

The bosonic fluctuation Lagrangian in static gauge is derived in
section~\ref{sec:ads3-bos-fluct} and reads
\begin{align}
  2 \mathcal{L}_{2B} &=
  \sinh^2\bar\rho \left( 1 + \frac{\omega^2\sinh^2\bar\rho}{(\bar\rho')^2} \right) \partial^a \tilde\phi \, \partial_a \tilde\phi
  + \left( 1 + \frac{\nu^2}{(\bar{\rho}')^2} \right) \partial^a \tilde\varphi \, \partial_a \tilde\varphi
  + \frac{2 \omega \nu \sinh^2\bar\rho}{(\bar\rho')^2} \partial^a \tilde\phi \, \partial_a \tilde\varphi 
  \nn \\ &\phantom{{}=}
  + \partial^a \tilde\beta_+ \, \partial_a \tilde\beta_+ + \partial^a \tilde\gamma_+ \, \partial_a \tilde\gamma_+
  + \partial^a \tilde\beta_- \, \partial_a \tilde\beta_- + \partial^a \tilde\gamma_- \, \partial_a \tilde\gamma_-
  + \partial^a \tilde{\psi} \, \partial_a \tilde{\psi}
  + \partial^a \tilde{U} \, \partial_a \tilde{U} 
  \nn \\ &\phantom{{}=} \label{LAdS3S3S3-2}
  + \nu_+^2 ( \tilde\beta_+^2 + \tilde\gamma_+^2 )
  + \nu_-^2 ( \tilde\beta_-^2 + \tilde\gamma_-^2 )
  \,.
\end{align}
The solution to the problem of finding \emph{exactly} the analytical partition function in cases of non-diagonal mass matrices like (\ref{LAdS3S3S3-2}) is not known at present.
An exact partition function can however be obtained for $\nu=0$, where the only non-trivial motion is the one in $\ads_3$. In this limit the coupling between the fluctuations $\tilde{\phi}$ and $\tilde{\varphi}$ disappears and the masses of the fluctuations in $\sphere^3 \times \sphere^3 \times \sphere^1$ vanish.
The bosonic contribution to the effective action then reads 
\begin{equation}
\tilde{\Gamma}_B=\frac{1}{2}\,\Big(\log\det\mathcal{O}_{\tilde\phi}+7\,\log\det\mathcal{O}_0\,\Big)
\end{equation}
where the operator $\mathcal{O}_{\tilde\phi}$ is the same as $\O_{\bar \phi}$ in (\ref{bosonoperatorsABJM}).

The full fermionic fluctuation Lagrangian derived in Appendix~\ref{app:fermionsADS3}) reads
\begin{align}
\mathcal L_{\textrm{GS}} =
2 i\rho' \bar\Psi\Big\{
&\Gamma^a\p_a - \frac{\kappa\, \omega\,\nu}{2 (\rho'^2+\nu^2)} \Gamma_{12}\bar{\Gamma} 
+ \frac{\rho'}{2} \Big( 
\Gamma^{012} - \sqrt{\alpha}  \Gamma^{345} - \sqrt{1-\alpha}  \Gamma^{678}
\Big)
\nn \\& \label{LGS-AdS3S3S3-2}
+ 
\frac{\sqrt{\rho'^2+\nu^2} - \rho'}{2}
\Big(
\Gamma^{012}
- \left(\sqrt{\alpha\delta} \Gamma^{34} + \sqrt{(1-\alpha)(1-\delta)} \Gamma^{67}\right) \bar{\Gamma}
\Big) 
\Big\} \Psi\,,
\end{align}
where we have chosen to fix the $\kappa$-symmetry by $\Psi^1= \Psi^2 \equiv \Psi$. 
As in the bosonic case,  an exact analytical derivation for the partition
function is only feasible in the $\nu=0$ case, where we obtain
\begin{equation}\label{fermads3}
  \mathcal L_{\textrm{GS}}= i \bar \Psi \bigg(\Gamma^a\p_a
    + \frac{\rho'}{2} \Big(\Gamma^{012} - \sqrt{\alpha} \Gamma^{345} - \sqrt{1-\alpha}  \Gamma^{678}\Big) 
  \bigg) \Psi \,,
\end{equation}
 which gives four massless fermions and four massive fermions, among which two have mass $\sqrt{\rho'^2-\rho''}$ and the remaining two have mass $\sqrt{\rho'^2+\rho''}$. 
The resulting fermionic contribution to the one-loop 2d effective action thus reads
\begin{gather} \label{fer-AdS3}
 \widetilde\Gamma_F = -\frac{1}{2} \Big( 2 \log \det \mathcal{O}_{\psi_+} + 2 \log \det \mathcal{O}_{\psi_-}+4\,\log\det\mathcal{O}_0 \Big) \,, 
 \\  
 \mathcal{O}_{\psi_\pm} = -\partial_\s^2+\Omega^2 +\mu^2_{\psi_\pm} \,, \qquad
 \mu^2_{\psi_\pm}= \pm \r'' + \r'^2 \,, \qquad
 \mathcal{O}_0=-\partial_\s^2+\Omega^2 \,.
 \label{fermops-AdS3}
\end{gather}

We note that for $\nu=0$ the fluctuation spectrum is independent of both
$\alpha$ and $\delta$, despite the presence of $\alpha$ in (\ref{fermads3}). In
particular we can take the limit $\alpha = \delta = 1$, which is further
discussed in section~\ref{AdS3-S3-T4}.

\subsection{Complete partition function, large and short string limits}
\label{sec:ads3-expansion}

From the frequencies in the last section we find that the one-loop correction to the energy of a folded string in $\ads_3$ is given by
\begin{equation}
  \label{general_E1_ads3}
  E_1= - \frac{1}{4 \pi\kappa} \int^{+\infty}_{-\infty} d\Omega \log \frac{\det \O_\psi^4}{\det \O_\phi \det\O_0^3}\,,
\end{equation}
where $\kappa$ is given in~\eqref{kappaomega} and we used that $\det\mathcal{O}_{\psi_+} = \det \mathcal{O}_{\psi_-}\equiv\det\mathcal{O}_\psi$. The
determinants appearing in \eqref{general_E1_ads3} are the same as in the cases of
$\ads_5$ and $\ads_4$. We will now evaluate this expression in the ``long
string'', or large spin, limit and in the ``short string'' limit, where the spin
is small. This is done in complete analog with the $\ads_4 \times \cp^3$
analysis in section~\ref{sec:long-and-short-ads4}.

\paragraph{Long string limit.}

In the long string limit ($k\rightarrow 1$ or $\eta\rightarrow 0$), we obtain
\begin{align}
  E_1 &= \frac{\kappa_0}{\kappa} \bigg [ 
  \left(c_{01} \kappa_0 +c_{00} + \frac{c_{0,-1}}{\kappa_0 }\right) + \left( c_{11} \kappa_0 + c_{10} + \frac{c_{1,-1}}{\kappa_0} \right) \eta 
  + \order(\eta^2)
\bigg] \,,
\end{align}
with
\begin{align}
  c_{01} &= -2 \log 2\,, &
  c_{00} &= \frac{4\log 2}{\pi} \,, &
  c_{0,-1} &= -\frac{1}{4}\,, \\ \nn
  c_{1,1} &= 0\,, &
  c_{10} &= -\frac{2\log 2}{ \pi}\,, &
  c_{1,-1} &= \frac{2\log 2}{\pi^2}\,.
\end{align}
Rewriting this in terms of the spin $\mathcal{\bar{S}}= 8 \pi\S$, we then get
\begin{align}\label{largeSads3}
  E_1 = 
  - \frac{2 \log 2}{\pi} \log\bar{\S} &{}+ \frac{4\log 2}{\pi} - \frac{\pi}{4\log\bar{\S}} \\ \nn
  &{}- \frac{1}{\bar{\S}} \left[ \frac{16\log 2}{\pi}\log\bar{\S} - \frac{24\log 2}{\pi} + \frac{\pi}{\log^2\bar{\S}} \right] + \order(1/\S^2)\,. 
\end{align}
Extracting the leading $\log S$ term we get
\begin{equation}
  \label{eq:ads3-scaling-func}
  E - S \approx f(\lambda) \log \frac{8\pi S}{\sqrt{\lambda}} \,, 
  \qquad
  f(\lambda) = \frac{\sqrt{\lambda}}{\pi} - \frac{2\log 2}{\pi} \,.
\end{equation}
This result is in perfect agreement with the one-loop part of the
$\ads_3\times\sphere^3\times\torus^4$ computation in Appendix~B
of~\cite{Iwashita:2011ha}. For a further discussion of this result see
Appendix~\ref{sec:ads3-scaling-limit}.

 In analogy with the analysis performed in section~\ref{sec:folded} one can check also in this case that reciprocity relations hold, evaluating the analog of (\ref{P1ABJM}) for the expansion (\ref{largeSads3}) above
\begin{equation}\label{P1ads3}
  \mathcal{P}_1=-\frac{2 \log2 \log \bar S}{\pi} - \frac{\pi}{4 \log\bar S}+\frac{4 \log 2}{\pi}+\order(\bar{\S}^{-2}) \,.
\end{equation}
and noticing its parity invariance under $\mathcal{S}\to-\mathcal{S}$.

\paragraph{Short string limit.}
Integrating out the zero modes we find the short string expansion to the
one-loop energy in~\eqref{general_E1_ads3} to be
\begin{equation}
E_1=  \sqrt{ 2 \S} \left(\frac{1}{2}-2 \log2 \right) +
\frac{1}{32} (2\S)^{3/2} ( -7+12 \log 2+6 \zeta(3)) +\O (\S) \,.
\end{equation}
Notice that in difference to the $\ads_5$ and $\ads_4$ cases, there is no
``non-analytical'' part of the type $E_1^{(\textrm{nan})}$ in (\ref{smallspin}). This is due to the vanishing of the 
total contribution from the zero modes, \textit{i.e.}, the eigenvalues with lowest energy, which is  in turn related to the absence of fluctuations 
transverse to the $\ads_3$ motion.


\subsection{The $\ads_3\times \sphere^3\times \torus^4$ limit}
\label{AdS3-S3-T4}

In the $\alpha \to 1$ limit one of the three-spheres in the $\ads_3 \times \sphere^3 \times \sphere^3 \times \sphere^1$ background becomes flat.\footnote{%
  The limit $\alpha \to 0$ is equivalent, with the other three-sphere blowing up.%
} %
Hence, this limit describes strings moving in $\ads_3 \times \sphere^3 \times \torus^4$, where the sphere now has the same radius as the anti-de Sitter space~\cite{Babichenko:2009dk}. In order to take this limit in the classical string solution considered above we first set $\nu_- = 0$ in~\eqref{classical_spinning}, corresponding to $\delta=1$ in the parametrization~\eqref{def-delta-parametrization}. We can then set $\alpha = 1$ in the fluctuation Lagrangians~\eqref{LAdS3S3S3-2} and~\eqref{LGS-AdS3S3S3-2}. We then obtain
\begin{align}
  2 \mathcal{L}_{2B} &=
  \sinh^2\bar\rho \left( 1 + \frac{\omega^2\sinh^2\bar\rho}{(\bar\rho')^2} \right) \partial^a \tilde\phi \, \partial_a \tilde\phi
  + \left( 1 + \frac{\nu^2}{(\bar{\rho}')^2} \right) \partial^a \tilde\varphi \, \partial_a \tilde\varphi
  + \frac{2 \omega \nu \sinh^2\bar\rho}{(\bar\rho')^2} \partial^a \tilde\phi \, \partial_a \tilde\varphi 
  \nn \\ &\phantom{{}=}
  + \partial^a \tilde\beta_+ \, \partial_a \tilde\beta_+ + \partial^a \tilde\gamma_+ \, \partial_a \tilde\gamma_+
  + \partial^a \tilde\beta_- \, \partial_a \tilde\beta_- + \partial^a \tilde\gamma_- \, \partial_a \tilde\gamma_-
  + \partial^a \tilde{\psi} \, \partial_a \tilde{\psi}
  + \partial^a \tilde{U} \, \partial_a \tilde{U} 
  \nn \\ &\phantom{{}=} \label{LAdS3S3T4}
  + \nu^2 ( \tilde\beta_+^2 + \tilde\gamma_+^2 )
  \,,
\end{align}
and
\begin{equation}
  \mathcal L_{\textrm{GS}} =
  i \bar\Psi\Big\{ 
  \Gamma^a\p_a - \frac{\kappa\, \omega\,\nu}{2 (\rho'^2+\nu^2)} \Gamma_{125} 
  + \frac{\sqrt{\rho'^2 + \nu^2}}{2} \Big( 
  \Gamma^{012} - \Gamma^{345}
  \Big)
  \Big\} \Psi\,.
\end{equation}
As noted above, the spectrum of fluctuations in the $\nu=0$ limit is independent
of $\alpha$. Hence, the one-loop energy of the folded string is the same in
$\ads_3 \times \sphere^3 \times \torus^4$ as in $\ads_3 \times \sphere^3 \times
\sphere^3 \times \sphere^1$. In particular the leading result for long strings
in~\eqref{eq:ads3-scaling-func} matches the calculation in Appendix~B
of~\cite{Iwashita:2011ha}.  For non-zero $\nu$ we note that the bosonic
fluctuations $\tilde{\beta}_-$ and $\tilde{\gamma}_-$ in~\eqref{LAdS3S3S3}
become massless when $\alpha=1$. These modes together with $\tilde{\psi}$ and
$\tilde{U}$ make up the four free directions on the torus.


 \bigskip

\section*{Acknowledgments }

We thank  B. Basso, M. Beccaria, B. Fiol, G. Grignani, L. Griguolo, F.  Levkovich-Maslyuk, D. Marmiroli, T. McLoughlin, C. Meneghelli, J. G. Russo, A.A.Tseytlin, K. Zarembo and in particular N. Drukker, L. Griguolo and D. Seminara for valuable discussions and comments on the draft. V.F.\ work is supported in part by the grant number FPA2010-20807,  the Consolider CPAN project and the [European Union] Seventh Framework Programme [FP7-People-2010-IRSES] under grant agreement n¡269217, in part by an Emmy Noether Programme funded by DFG. V.G.M.P.\ acknowledges  NORDITA and the  Swedish Research Council for funding under the contract 623-2011-1186. O.O.S.\ acknowledges support from the Netherlands Organization for Scientific Reasearch (NWO) under the VICI grant 680-47-602.


\appendix

\section{Notation and $\Gamma$ matrices representation}
\label{app:gamma_matrices}

We denote with  $A,B,\dots =0,\dots, 9$ the  10d  flat indices, with $M, N,\dots$ the 10d curved ones, and with $a,b =0,1$ the world-sheet indices. Finally, the indices $I,J,K$ label the two kinds of spinors $\theta^1, \theta^2$, or  $\Psi^1, \Psi^2$.

For the type IIA string in $\ads_4 \times \cp^3$, the spinors are defined as $\theta=\theta^1+\theta^2$ with
$
\Gamma_{11}\theta^1=\theta^1
$
and
$
\Gamma_{11}\theta^2=-\theta^2\,, 
$
while for the type IIB string in $\ads_3/\mbox{CFT}_2$ they have the same chirality $\theta^I = \Gamma_{11}\theta^I$, $I=1,2$. 
The complex conjugate $\bar\theta$ is always defined as $\bar\theta^I = \theta^I \Gamma^0$. 
Finally, we use the following representation for the 10d Dirac matrices
\begin{equation}
\Gamma_0 = -i \sigma_2 \otimes I_{16} 
\qquad\quad
 \Gamma_{11} = \sigma_3 \times I_{16} 
\qquad\quad
\Gamma_a = \sigma_1 \otimes \hat{\gamma}  
\,, \quad a=1,\dotsc,9 \,. 
\end{equation}


\section{Fluctuations in  $\ads_4 \times\cp^3$}
\label{app:AdS4xCP3}
 
The Type IIA background $\ads_4 \times\cp^3$ is defined via
\begin{gather}
  \label{total_metric}
ds^2=R^2\left ( ds^2_{\ads}+ 4 ds^2_{\cp^3}\right) \,,
  \qquad
  e^\phi= \frac{2 R}{ k} \,,
  \qquad
  R^2 \equiv {\widetilde{R}^3\over 4 k}= \frac{k^2}{4} e^{2\phi} \,,
  \\
  F_2 = 2\,k\,J_{\cp^3} \,,
  \qquad
  F_4=\frac{3}{8} R^3\, \vol(\ads_4) \,,
\end{gather}
where $R$ is the $\ads_4$ radius, $\phi$ the dilaton, $k$  results from the compactification of the original M-theory on $\ads_4\times \sphere^7/Z_k$ background and coincides in the dual theory with the super Chern-Simons matter theory level number~\cite{Aharony:2008ug}. Above, $F_2$, $F_4$ are the 2-form and 4-form field strengths with  $J_{\cp^3}$ the K\"ahler form on $\cp^3$. The curvature radius $R$ is related to the 't Hooft coupling $\lambda$ of the dual $\mathcal{N}=6$ superconformal Chern-Simons theory (realized in the limit of $k$ and $N$ large with  their ratio fixed)  
\begin{equation}\label{lambdaABJM}
R^2=2^{5/2}\pi\,\sqrt{\lambda} \,, \qquad\lambda=\frac{N}{k}\,.
\end{equation}

To describe the background (\ref{total_metric}) we use global coordinates for $\ads_4$
\begin{equation}
\label{metricAdS4}
ds^2_{\ads}= -\cosh^2\rho\,dt^2 + d\rho^2 + \sinh^2\rho (d\phi^2 + \sin^2\phi\,d\varphi^2)
\end{equation}
and a parametrization~\cite{Hoxha:2000jf} in terms of six coordinates $(\psi_1,\psi_2,\psi_3,\tau_1,\tau_2,\tau_3)$ for $\cp^3$
\begin{equation}\label{metricCP3}
  \begin{aligned}
    ds^2_{\cp^3} &= d\psi_1^2 +\sin^2\psi_1 d\,\psi_2^2 + \sin^2\psi_1 \cos^2\psi_1\left( d\tau_1 + \sin^2\psi_2 (d\tau_2 + \sin^2\psi_3\,d\tau_3)\right)^2 \\
    &+\sin^2\psi_1 \sin^2\psi_2 \left( d\psi_3^2 + \cos^2\psi_2 (d\tau_2 + \sin^2\psi_3\,d\tau_3)^2 + \sin^2\psi_3 \cos^2\psi_3\,d\tau_3^2\right)\,.
  \end{aligned}
\end{equation}
Alternatively, one might use three inhomogeneous complex coordinates 
\begin{align}\nonumber
  w_1 &= { \tan\psi_1 \tan \psi_2 \tan\psi_3 \over \sqrt{\left(1+\tan^2\psi_2\right)\left(1+\tan^2\psi_3\right)} }e^{i \left(\tau_1+\tau_2+\tau_3\right)}\,,\\ \label{coords_w}
  w_2 &= { \tan\psi_1 \tan \psi_2 \over \sqrt{\left(1+\tan^2\psi_2\right)\left(1+\tan^2\psi_3\right)} }e^{i \left(\tau_1+\tau_2\right)}\,,\\ \nn
  w_3 &= { \tan\psi_1 \over \sqrt{1+\tan^2\psi_2}} e^{i \tau_1}\,,
\end{align}
in terms of which the metric
\begin{equation}
ds^2_{\cp^3}=\frac{d\bar{w}_i\,dw_i}{1+|w|^2}-\frac{d\bar{w}_i\,dw_i\,d\bar{w}_j\,dw_j}{(1+|w|^2)^2} \,, \qquad |w|^2=\bar{w}_k \,w_k \,,
\end{equation}
gives back the Fubini-Study metric \eqref{metricCP3}. 

Another useful parametrization of $\cp^3$ is the one describing its embedding  in $\C^4$
\begin{align}\label{embCP3}
  z_1 &= \cos\frac{\alpha}{2}\cos\frac{\vartheta_1}{2}e^{i\frac{\varphi_1}{2}+i\frac{\chi_1}{4}} \,,\\
  z_2 &= \cos\frac{\alpha}{2}\sin\frac{\vartheta_1}{2}e^{-i\frac{\varphi_1}{2}+i\frac{\chi_1}{4}} \,,\\
  z_3 &= \sin\frac{\alpha}{2}\cos\frac{\vartheta_2}{2}e^{i\frac{\varphi_2}{2}-i\frac{\chi_1}{4}} \,,\\
  z_4 &= \sin\frac{\alpha}{2}\sin\frac{\vartheta_2}{2}e^{-i\frac{\varphi_2}{2}-i\frac{\chi_1}{4}} \,,
\end{align}
with $0\leq\alpha,\vartheta_1,\vartheta_2<\pi$, $0\leq\varphi_1,\varphi_2<2\pi$, $0\leq\chi<4\pi$. This leads to the Fubini-Study metric
\begin{align}\label{CP32}
ds^2_{\cp^3} &=d\alpha^2+\cos^2\frac{\alpha}{2}\,(d\vartheta_1^2+\sin^2\vartheta_1\,d\varphi_1)+\sin^2\frac{\alpha}{2}\,(d\vartheta_2^2+\sin^2\vartheta_2\,d\varphi_2)+\\\nonumber
 &\qquad\qquad\qquad\qquad\qquad+\sin^2\frac{\alpha}{2}\cos^2\frac{\alpha}{2}\,(d\chi+\cos\vartheta_1\,d\varphi_1-\cos\vartheta_2\,d\varphi_2)^2\,.
\end{align}
From the inhomogeneous coordinates (\ref{coords_w}) one moves to the homogeneous coordinates (\ref{embCP3}) via the map $w_i =z_i/z_4$, $i=1,2,3$ and the condition $|z|^2=1$.  
 
In this background we study quadratic fluctuations over  classical string configurations which solve the equation of motion for the bosonic action
 \begin{equation}
\label{bosonic_action}
I_{\rm B} = - {R^2 \over 4\pi \alpha'} \int d\tau d\sigma \sqrt{-h} h^{ab} \left (G^{\ads}_{MN} \p_a X^M \p_b X^N + 4\,G^{\,\cp}_{MN} \p_a X^M \p_b X^N \right)\,.
\end{equation}
For this purpose we also need the quadratic term in the fermionic part of the Green-Schwarz action
\begin{equation}
\label{def_Lfermions}
\CL_{F}=i \left( \sqrt{-h}h^{ab} \delta^{IJ} -\epsilon^{ab} s^{IJ} \right) \bar\theta^I \rho_a \, D_b^{JK} \theta^K\,,
\end{equation}
where the form of the covariant derivative is restricted by the background RR fluxes 
\begin{align}\label{def_covariant_der}
  D_b^{JK} \theta^K &=
  \delta^{JK} \left ( \p_b + \quarter \p_b X^N \omega_N^{BC} \Gamma_{BC} \right) \theta^K
  \\ \nn
  &\quad+ {1\over 8} e^\phi \left( \half F_{MN} \Gamma^{MN} (i\sigma_2)^{JK} 
    +{1\over 4!} F_{MNPQ} \Gamma^{MNPQ} (\sigma_1)^{JK}\right) \rho_b\,  \theta^K\,,
\end{align}
with $\rho_a = \p_a X^M E^A_M \Gamma_A$ and $s^{IJ}= \text{diag} (1,-1)$,  where we used already that $H_{(3)}$,  the three-form potential for $F_4$, will  not play a role in our calculations.

\subsection{Antiparallel lines}
\label{app:lines}

The classical solution and the formal expression for bosonic fluctuations over it were considered for the $\ads_5\times \sphere^5$ background in~\cite{Drukker:2000ep}. As explained in 
section~\ref{sec:lines}, since the classical solution is here trivial  in $\cp^3$ the bosonic spectrum is a simple truncation (one less transverse fluctuation) of the one written down in~\cite{Drukker:2000ep}, and the contribution to the effective action reads
\begin{equation}
  \tilde{\Gamma}_B=\frac{1}{2}\,\Big(\log\det\mathcal{O}_\phi+\log\det\mathcal{O}_\theta+6\,\log\det\mathcal{O}_0\,\Big)\,.
\end{equation}
The fermionic fluctuations are obtained by substituting in (\ref{def_Lfermions})-(\ref{def_covariant_der}) the relevant classical ansatz. If the  Wilson lines are extended in the $x^0\equiv\tau$ direction and  located at $x^1\equiv\sigma=\pm L$, the ansatz for the corresponding minimal surface is given by 
$
y=y(\sigma), 
$ 
where $y$ is the radial direction in AdS.  The derivation of the resulting fermionic Lagrangian works much the same as in the case of the folded string solution~\cite{McLoughlin:2008ms,McLoughlin:2008he} reviewed below in section~\ref{sec:folded}, leading to
\begin{equation}\label{FermionsLinesABJM}
\mathcal{L}_F 
= 2 i y^4  \bar\Psi  \left\{
 \eta^{ab}\Gamma_a \p_b + \tilde{\Gamma}\right\} \Psi
 \,, \qquad
\tilde \Gamma = {1\over 4} \left(  - \Gamma_{47}+\Gamma_{58}-\Gamma_{69} -3 \Gamma_{0123}\right)\,.
\end{equation}
One notices that it is always possible to further simplify the problem, bringing the $32\times 32$ matrix $\tilde\Gamma$  above into a $2\times 2$ block-diagonal form in block with 4 vanishing entries and the remaining 12 consisting of $\sigma_3$ matrices, where $\sigma_3$ is the third Pauli matrix.\footnote{The relevant rotation is built with eigenvectors of the matrix $\Gamma_0\Gamma_1$, which commutes with $\tilde\Gamma$.} The same rotation makes block-diagonal the kinetic term in (\ref{FermionsLinesABJM}), with  
$\Gamma_0={\rm I}_{16}\times \sigma_1, \Gamma_1={\rm I}_{16}\times \sigma_2$ where $\sigma_1,\sigma_2$ are the remaining Pauli matrices. 
Thus, one ends with 8 species of 2-d Majorana fermions  two of which are massless and cancelled by the massless bosons, and the final result reads as in  (\ref{ZlinesABJM}). The rest of the analysis goes as in~\cite{Forini:2010ek}, based on which it is not difficult to check that the final result is given  in terms of an analytical and a part which  is numerically integrable with arbitrary precision, where the latter is defined via
\begin{align}\no
\mathcal{I}^{\textrm{num}} &= \int_0^{\pi} \frac{d\a}{2\cos^2\textstyle{\frac{\alpha}{2}}}
\Big[\log\frac{(1+e^{-2 \bar{x}_f})^2}{1-e^{-2\bar{x}_f}}-\frac{\sqrt{2}}{2} \sin\textstyle{\frac{\a}{2}}
\Big(\log(1-e^{-2\,x_{1{\textrm{inf}}}})+\cos^2\frac{\a}{2}\log\sin x_{1{\textrm{zero}}}\Big)\Big] \,, \\\no
\bar x_f &= \frac{1}{2}\Big(\frac{\pi}{2\,\KK}+\KK\Big)\,F[\alpha]-\KK\,E[\a]+\frac{\KK}{2}\,\sin\a\,\sqrt{1+\tan^4\frac{\a}{2}} \,, \\
\bar x_{1{\textrm{zero}}} &= \Big(\frac{\pi}{2\,\KK}-\KK\Big)\,F[\textstyle{\frac{\alpha}{2}}]+2\,\KK\,E\textstyle{\frac{\alpha}{2}}] \,, \\\no
\bar x_{1{\textrm{inf}}} &= \Big(\frac{\pi}{2\,\KK}+\KK\Big)\,F[\textstyle{\frac{\alpha}{2}}]-2\KK\,E[\textstyle{\frac{\alpha}{2}}]+ \KK \tan\frac{\alpha}{2}\, \sqrt{3 + \cos\a} \,,
\end{align}
which gives
\begin{equation}\label{numericsABJM}
\mathcal{I}^{\textrm{num}}=2.118816702312941142665389260961407916565757084503648488839669\,.
\end{equation}

\subsection{Generalized cusp}
\label{app:cusp}

The classical string solution corresponding to the Wilson loop introduced in section~\ref{sec:cusp} coincides (up to the mapping $\lambda\to\bar\lambda$) of the one for the $\ads_5\times \sphere^5$ background  in~\cite{Drukker:2011za}, to which we refer the reader for more details. Below we report the expressions which are strictly necessary to define the relevant formulas  here used.

The Nambu-Goto action reads 
\begin{equation}
\mathcal{S}_{NG}=\frac{\sqrt{\bar{\lambda}}}{2\pi}\int dt\,d\varphi
\cosh\rho\sqrt{\sinh^2\rho+(\partial_\varphi\rho)^2+(\partial_\varphi\vartheta)^2}\,,
\label{NG}
\end{equation}
and one finds the same set of conserved charges
\begin{equation}
E=-\frac{\sinh^2\rho\cosh\rho}{\sqrt{\sinh^2\rho+(\partial_\varphi\rho)^2+(\partial_\varphi\vartheta)^2}}\,,
\qquad
J=\frac{\partial_\varphi\vartheta\cosh\rho}{\sqrt{\sinh^2\rho+(\partial_\varphi\rho)^2+(\partial_\varphi\vartheta)^2}}\,.
\label{EJ}
\end{equation}
The BPS condition is still given by $E=\pm J$, or $\phi=\pm\theta$~\footnote{One might refer to the vanishing of the total Lagrangean, see~\cite{Drukker:2007qr} Appendix C.2, which does not change here. A study of the BPS condition on the gauge theory side, which is also useful for nonperturbative computations is carried out in~\cite{Griguolo:2012iq}.}, and one conveniently introduces the two parameters 
\begin{equation}
\label{def_pq}
q=-\frac{J}{E}\,, \qquad p=\frac{1}{E} \,,
\end{equation}
which are related to $\phi$ and  $\theta$ via the transcendental equations~\cite{Drukker:2011za}
\begin{equation}\label{phitheta}
\phi=\pi-2\frac{p^2}{b\sqrt{b^4+p^2}}
\left[\Pi\big({\textstyle\frac{b^4}{b^4+p^2}}|k^2\big)-\KK(k^2)\right] \,, \qquad
\theta=\frac{2b\,q}{\sqrt{b^4+p^2}}\KK(k^2)\,,
 \end{equation}
where $b,k$ are in turn related to $p,q$ via
\begin{equation}\label{pq}
  p^2=\frac{b^4(1-k^2)}{b^2+k^2}\,,\qquad
  q^2=\frac{b^2(1-2k^2-k^2b^2)}{b^2+k^2}\,.
\end{equation}
The classical equations of motion for $\rho$ and $\vartheta$ appearing in (\ref{NG}) read then
\begin{equation}\label{eom}
  (\partial_\varphi\rho)^2=p^2\cosh^2\rho\sinh^4\rho-q^2\sinh^4\rho-\sinh^2\rho\,,\qquad
  \partial_\varphi\vartheta= q\, \sinh^2\rho\,,
\end{equation}
and the result for the classical action  is
\begin{equation}
  \mathcal{S}_\text{cl}=\frac{T\sqrt{\bar{\lambda}}}{2\pi}\frac{2\sqrt{b^4+p^2}}{b\,p}
  \left[\frac{b\,p\,\sinh\rho_0}{\sqrt{b^4+p^2}}+\frac{(b^2+1)p^2}{b^4+p^2} \KK(k^2)-\EE(k^2)\right],
  \label{classical-action}
\end{equation}
where $T$ is a cutoff on the $t$ integral,
$E$ denotes an elliptic integral of the second kind, and $\rho_0$ is a cutoff at large $\rho$ to cure the standard linear divergence 
for two lines along the boundary.  As usual, one can cancel this divergence by
a boundary term, remaining with the second and third term within brackets in (\ref{classical-action}).

It is useful to redefine the world-sheet coordinates $(t,\varphi)$ in terms of elliptic ones
\begin{equation}\label{rescaling}
\sigma=\frac{\sqrt{b^4+p^2}}{b\,q}\,\vartheta=F(\arcsin\xi|k^2)-\KK(k^2)\,,
\qquad
\tau=\frac{\sqrt{b^4+p^2}}{b\,p}\,t\,,
\end{equation}
where the range of the new coordinates is
\begin{equation}\label{range}
-\KK<\sigma<\KK\,,\qquad-\infty<\tau<\infty\,.
\end{equation}
The relation between $\xi$ and $\varphi$ in (\ref{rescaling}) is given in terms of incomplete 
elliptic integrals of the first and third kind $F$ and $\Pi$~\cite{Drukker:2011za} and one also finds 
\begin{equation}\label{rhocn}
  \xi=\sn(\sigma+\KK)=\frac{\cn(\sigma)}{\dn(\sigma)}\,,
  \qquad
  \cosh^2\rho=\frac{1+b^2}{b^2\cn^2(\sigma)}~.
\end{equation}
The nice effect of the rescaling (\ref{rescaling}) is to put the induced metric in a conformally flat form
\begin{equation}
ds^2_\text{ind}
= \frac{1-k^2}{\cn^2(\sigma)}\big[{-}d\tau^2+d\sigma^2\big].
\label{ind-metric}
\end{equation}
The 2--dimensional scalar curvature reads
\begin{equation}
R^{(2)}=-2\left(1+ \frac{k^2(1 + b^2)^2 }{b^4 (1 - k^2)\cosh^4\rho}\right)
=-2\left(1+\frac{k^2}{1-k^2}\cn^4(\sigma)\right).
\label{curvature}
\end{equation}
It is useful to rewrite the equations of motion (\ref{eom}) for $\rho,\vartheta$ and add the one for $\varphi$~\cite{Drukker:2011za}
\begin{equation}\label{eqthphs}
~~ \mathllap{\rho'^2=}\frac{ (b^2\sinh^2\rho - 1) (b^2 + p^2\sinh^2\rho)}{(b^4+p^2)\sinh^2\rho} \,, \quad
  \vartheta'^2=\frac{p^2\,(b^2+1) -b^4}{b^4+p^2} \,, \quad
  \varphi'^2=\frac{b^2}{(b^4+p^2)\sinh^4\rho} \,.
\end{equation}

We also report here the expansion of the parameters $\phi,\theta$ in the relevant limit  $\phi=\theta=0$, which is a $p\to\infty$ limit.
\begin{align}\label{phithetapinfinity}
\phi&=\frac{\pi}{p}+\frac{\pi(3q^2-5)}{4p^3}+\frac{3\pi(15q^4-70q^2+63)}{64p^5} \\\nn
&\qquad\qquad\qquad\qquad\qquad+\frac{5\pi(7q^2(5q^4-45q^2+99)-429)}{256p^7}+\order(p^{-9})\,,
\\
\theta&=\frac{\pi q}{p}+\frac{\pi q(q^2-3)}{4p^3}+\frac{3\pi q(3q^4-30q^2+35)}{64p^5} \\\nn
&\qquad\qquad\qquad\qquad\qquad+\frac{5\pi q
(5q^2(q^4-21q^2+63)-231)}{256p^7}+\order(p^{-9})\,.
\end{align}

\subsubsection{Fluctuations}

In carrying out the study of fluctuations, it turns out to be easier to use different parametrizations of the classical ansatz (\ref{ansatz}) for the ``jump'' in $\cp^3$. In particular, while for the study of the fermionic spectrum we will be using the six coordinates in (\ref{metricCP3}) and fluctuate over the ansatz
\begin{equation}\label{ansatzfermions}
   \psi_1 = {\pi \over 4} \,, \qquad
\psi_2 = {\pi \over 2}   \,, \qquad
\psi_3= {\pi \over 2}   \,, \qquad
\tau_1= \tau(\sigma)   \,, \qquad
 \tau_3= \tau(\sigma)   \,, \qquad
\tau_2=0\,,
\end{equation}
in the case of bosonic fluctuations we will use the parametrization (\ref{coords_w}), in terms of which 
the ansatz (\ref{ansatzfermions}) reads
\begin{equation}\label{ansatzw}
\psi_1 = {\pi \over 4}\, \qquad
\psi_2 = {\pi \over 2}   \,, \qquad
\psi_3= {\pi \over 2}   \,. \qquad\,
w_1 = e^{ 2\,i\,\tau(\sigma)}\,,~~~~w_2 = w_3 =0\,.
\end{equation}
As an example, we could use the ans\"atze above (clearly equivalent to each other), as well as the one in  (\ref{ansatz}), in the proposal~\cite{Berenstein:2008dc,Rey:2008bh,Drukker:2009hy,Lee:2010hk} for the SU(4) tensor $M^I_{~J}$  representing the coupling $\sim |\dot{x}^\mu|\,M^I_J\,C_I\,\bar C^J$ to the bilinear scalars in the ABJM Wilson loop
\begin{equation}\label{soojong}
  M^I_{~J}=\delta^I_{~J} -2\frac{\bar{z}_J\,z^I}{|z|^2}\,,
\end{equation}
 written in terms of four internal $SU(4)$ coordinates $z^I$ which are  in a one-to-one correspondence with the scalars $C^I$ in ABJM theory. It is natural to identify the $z^I$ above with the coordinates embedding $\cp^3$ in $\C^4$ (\ref{embCP3}). From the ansatz (\ref{ansatz})  it then follows\footnote{Along the boundary, the $\varphi$-dependence of $\theta_1$ is the $s$-dependence along the loop.}  
\begin{equation}\label{jump}
  \widetilde{M}^I_{~J}=\begin{pmatrix}-\cos\theta_1&-\sin\theta_1&0&0\\-\sin\theta_1&\cos\theta_1&0&0\\
    0&0&1&0\\
    0&0&0&1\end{pmatrix},
\end{equation}
which evaluated at the extrema of the $\theta_1$-interval, \emph{i.e.}, at the
boundary of $\ads$, should give a representation of the ``jump'' in the internal
space for the Wilson loop, at least for the scalar coupling.\footnote{%
  In the case of the ans\"atze (\ref{ansatzw}), we can set $z_a = \left(z_4\,,
    w_1 z_4\,, w_2 z_4\,, w_3 z_4 \,\right)$ with $z_4= {e^{-i \tau} \over
    \sqrt{1+\tan^2\psi_1}}$, from which it follows for the coupling
  (\ref{soojong})
  \begin{equation}\label{jump2}
    \widetilde{M}^I_{~J}=\begin{pmatrix}
      0 &-e^{-i \tau} &0&0\\
      -e^{i \tau} &0&0&0\\
      0&0&-1&0\\
      0&0&0&-1\end{pmatrix}.
  \end{equation}
  It is always possible to bring (\ref{jump2}) above in the form (\ref{jump})
  via a $\grSU(4)$ transformation.%
} %
The matrix (\ref{soojong}) as well as (\ref{jump}) above are locally
SU(3)-invariant matrices. Explicitly, on each of the lines forming the Wilson
loop, they can always (separately) be brought to the form $\bar{M}={\rm
  diag}(-1,1,1,1)$ typical of the half-BPS Wilson line operator corresponding to
the most symmetric string configuration.
\subsubsection*{Bosonic fluctuations}

In evaluating the bosonic fluctuations for the generalized cusp we expand the Nambu-Goto action  around the classical solution via the following variations
\begin{equation}
\rho= \rho(\sigma)+\delta\rho \,, 
\qquad
\varphi = \varphi(\sigma) +\delta \varphi\,,\qquad \phi=\frac{\pi}{2}+\delta\phi
\end{equation}
in $\ads_4$ and fluctuations over the ansatz (\ref{ansatzw}) in $\cp^3$. Closely following~\cite{Drukker:2011za}, we use a static gauge where longitudinal fluctuations are set to zero, leaving the system only with transverse physical degrees of freedom. The fluctuations parallel to the world-sheet are $\delta t$ (set to zero from the beginning) and a linear combination of $\delta\rho\,, \delta \varphi\,\delta\vartheta$, where here $\vartheta\equiv \tau_1$.
Since the classical Lagrangian for the solution \eqref{ansatz} is the same as in the $\ads_5\times \sphere^5$ case, the rotation that individuates the directions normal  to the world-sheet will stay the same.
Thus, we can construct the transverse fluctuations out of  $\delta\rho\,, \delta \varphi\,,\delta \vartheta$ as in~\cite{Drukker:2011za}
\begin{equation}
  \mathllap{\delta \xi_7} =\frac{\vartheta'(\sinh^2\rho\,\varphi'\,\delta\varphi+\rho'\,\delta\rho) -(\rho'^2+\sinh^2\rho\,\varphi'^2)\delta\vartheta}
  {\sqrt{(\rho'^2+\sinh^2\rho\,\varphi'^2)(\vartheta'^2+\rho'^2+\sinh^2\rho\,\varphi'^2)}}\,, \qquad
  \delta \xi_8= \frac{\sinh\rho\,(\rho'\,\delta\varphi-\varphi'\,\delta\rho)}{\sqrt{\rho'^2+\sinh^2\rho\,\varphi'^2}}\mathrlap{\,.}
\end{equation}
We expect that the only difference with the $\ads_5\times \sphere^5$ bosonic spectrum results in  one less transverse  $\ads$ fluctuation and a splitting among ``heavy'' and ``light'' excitations in the transverse directions of $\cp^3$. Indeed, the resulting  Lagrangian is
\begin{equation}\label{bosoniclagr}
\mathcal L_{B}= \frac{1}{2}\sqrt{g}\Big[g^{ab}\,\partial_a \zeta_P\,\partial_b \zeta_P
+A(\zeta_8\partial_\sigma\zeta_7-\zeta_7\partial_\sigma\zeta_8)
+M_{PQ}\zeta_P\zeta_Q\Big],
\end{equation}
where
$\zeta_1= \delta \phi\sinh\rho$ is the transverse direction in $\ads_4$\,, $\zeta_2 =\psi_1$ is the heavy transverse direction in $\cp^3$, and $\zeta_s$, $s=3,\dots, 6$ the remaining transverse light excitations in the compact space. The corresponding masses are
\begin{align}\label{bosonicmassesfirst}
M_{11} &= {b^4 -b^2 p^2 -p^2\over b^2 p^2 \cosh^2\rho}+2 \,,\qquad 
M_{22}=  {b^4 -b^2 p^2 -p^2\over b^2 p^2 \cosh^2\rho}\,, \\
M_{ss} &=  {b^4 -b^2 p^2 -p^2\over 4 \,b^2 p^2 \cosh^2\rho}\,, \quad s=3,4,5,6 \,,
\\ 
M_{77} &= \frac{b^4-b^2p^2-p^2}{b^2p^2\cosh^2\rho}
-\frac{2(b^2+1)(b^2-p^2)}{b^2p^2\cosh^4\rho}
+b^2\frac{b^4+2b^2p^2\sinh^2\rho+b^2p^2-p^2}
{\cosh^2\rho\left(b^4+b^2p^2\sinh^2\rho-p^2\right)^2}
\\ \nn
M_{88} &= \frac{b^4-b^2p^2-p^2}{b^2p^2\cosh^2\rho}+2
-\frac{3b^2}{\cosh^2\rho(b^4+b^2p^2\sinh^2\rho-p^2)}
+\frac{b^4p^2}{\left(b^4+b^2p^2\sinh^2\rho-p^2\right)^2}\,,
\\ 
M_{78} &= M_{87} = \frac{2\sqrt{-b^4+b^2p^2+p^2}\sqrt{b^2\sinh^2\rho-1}\sqrt{b^2+p^2\sinh^2\rho}}
{p(b^4+b^2p^2\sinh^2\rho-p^2)\cosh^3\rho}\,,
\\ \label{bosonicmasseslast}
A &=\frac{2\sqrt{b^4+p^2}\sqrt{-b^4+b^2p^2+p^2}}{p\left(b^4+b^2p^2\sinh^2\rho-p^2\right)\cosh^2\rho}~.
\end{align}
They match exactly the $\ads_5 \times \sphere^5$ case~\cite{Drukker:2011za}, apart from the factor 4 in the light modes $M_{ss}$.

\subsubsection*{Fermionic fluctuations}

Fermionic fluctuations are computed from the Green-Schwarz Lagrangian \eqref{def_Lfermions} with the covariant derivative as in \eqref{def_covariant_der}. On the classical ansatz \eqref{ansatz} the Gamma matrices pulled-back onto the world-sheet $\rho_a = \p_a X^M E_M^A \Gamma_A$ read
\begin{equation}
\rho_0 = 
 \kappa r_0 \Gamma_0\,, \qquad
\rho_1 =
\rho' \Gamma_1 +\varphi' r_1 \Gamma_3 +\alpha'_s \Gamma_7\,,
\qquad r_1 \equiv \sinh\rho \,, \quad r_0 \equiv \cosh\rho\,,
\end{equation}
where we rescaled the isometry in $\cp^3$ according to $\alpha_s = \left(\tau_1+\tau_2+\tau_3\right)/4$ in order to make contact with the classical solution in $\ads_5 \times \sphere^5$, and  introduced the constant factor $\kappa$ to rescale the time coordinate for later convenience.
 
 The effect of the local spinor rotation 
$\theta^I = S\, \Psi^I$, where 
\begin{gather}\label{xizetacusp}
  S= \left(\cos{{\xi\over 2}}+\sin{{\xi\over 2}} \Gamma_{13}\right) \left(\cos{{\zeta\over 2}}+\sin{{\zeta\over 2}} \Gamma_{17}\right) \\\nonumber
  \cos\xi=\left( \frac{  \left( b^2 +p^2 \sinh^2 \rho \right) \left( b^2\sinh^2\rho -1\right) }{ \sinh^2\rho \left( b^4 +p^2 b^2 \sinh^2\rho -p^2\right) }\right)^{1/2} \!\!\!\!\!\!\!\!,
  \qquad
  \cos\zeta = \frac{\left(  b^4 +p^2 b^2 \sinh^2\rho -p^2 \right)^{1/2}}{\cosh\rho \sqrt{b^2 p^2}}\,,
\end{gather}
is to bring  the pulled-back Gamma matrices into the two-dimensional  form $\{\tau_i,\tau_j\}=g_{ij}$ 
\begin{equation}
\tau_0 \equiv S^{-1} \rho_0 S=a \Gamma_0\,, \qquad
\tau_1 \equiv S^{-1} \rho_1 S= a \Gamma_1\,, 
\qquad \text{with}\quad a = \sqrt{\frac{b^2 p^2 \cosh^2\rho}{b^4 +p^2 }}\,. 
\end{equation}
Above, $a$  is the fourth-root of the determinant in the conformally flat induced metric (\ref{ind-metric}),  and can also be written, as from (\ref{rhocn}), as $a=\frac{\sqrt{1-k^2}}{\cn\left(\sigma \left|k^2\right.\right)}$. 
The rotated covariant derivatives are
\begin{align}
  \nabla_0^{IJ} &\equiv S^{-1} \mathcal D_0^{IJ} S= \delta^{IJ} \left( \p_0 + {a r_1 \over 2 r_0} \left(\cos\xi \sin\zeta \Gamma_{07} +\cos\xi \cos\zeta \Gamma_{01}\right)+  {a r_1 \over 2 r_0}\sin\xi \Gamma_{03}\right) \,,
  \nn \\
  \nabla_1^{IJ} &\equiv S^{-1} \mathcal D_1^{IJ} S= \delta^{IJ}
  \left( {\zeta' \over 2} \Gamma_{17} +\half \left( \xi' -\varphi' r_0\right) \left( \Gamma_{13} \cos\zeta -\Gamma_{37} \sin\zeta\right)\right)\,.
\end{align}
The fluxes contribute as
\begin{align}
\hat{\mathcal{F}}^{JK} &\equiv S^{-1} \mathcal F^{JK} S =  {e^\phi \over 8} S^{-1} \left( \half \Gamma^{MN} F_{MN} \left(i \sigma_2\right)^{JK} +{1\over 4!} \Gamma^{MNPQ} F_{MNPQ}\sigma_1^{JK} \right)S
\\ \nn
& = {k e^\phi \over 16} \left(  \left(i \sigma_2\right)^{JK} \left( \Gamma_{69}-\Gamma_{58}+\sin\zeta \Gamma_{13} +\cos\zeta \Gamma_{47} \right)
- 3 \sigma_1^{JK} \left( \cos\zeta\, \Gamma_{0123} +\sin\zeta \,\Gamma_{0237}\right)\right)\,.
\end{align}
Collecting the kinetic terms and the fluxes we have\footnote{Terms of the type $\bar\Psi\,\Gamma_a\,\Psi$ were written for completeness, since they vanish and can thus be ignored. In \eqref{final_2_fermion_cusp}  we set the $AdS_4$ radius to 1.}
\begin{align}
\label{final_fermion_cusp}
\mathcal L_{F}  &=  i \left( \sqrt{-h}h^{ab}\delta^{IJ}-\epsilon^{ab} s^{IJ}\right)\bar\Psi^I \left( \tau_a \nabla_b^{JK} +
 \tau_a \mathcal F^{JK}\tau_b \right) \Psi^K
 =\\ \nn
&= 2a\,i \, \bar\Psi \Big\{
\eta^{ab} \Gamma_a \p_b \CP
+{a r_1\over r_0 } \cos\xi \cos \zeta \Gamma_1\, \CP 
- \left( {a r_1\over r_0 }  \sin\xi+ \left( \xi' -r_0 \varphi' \right) \cos\zeta \right) \Gamma_3 \CP
 \nn \\
&\qquad\qquad\qquad-  \left({a r_1\over r_0 } \cos\xi \sin\zeta +\zeta' \right) \Gamma_7\, \CP
 - \sin\zeta  \left(\xi' -r_0 \varphi' \right) \Gamma_{137} \CP
\nn \\\label{final_2_fermion_cusp}
&\qquad\qquad\qquad\qquad\qquad- {a \over 4}  \left( \left(\Gamma_{69}-\Gamma_{58} +\cos\zeta\,\Gamma_{47}\right)\Gamma_{11} +3 \cos\zeta\, \Gamma_{0123} \right) \CP \Big\}
\Psi\,,
\end{align}
where as in~\cite{McLoughlin:2008he} we have reorganized the spinors via a suitable projector operator 
\begin{equation}
\label{def_P}
\mathcal P\equiv\frac{1}{2}\left( 1+\Gamma_0\Gamma_1\Gamma_{11}\right)~,
\end{equation}
and a linear combination~$\Psi=\Psi^1+\Psi^2$ of the two Type IIA spinors with opposite chirality has been introduced. A natural gauge choice to fix the $\kappa$-symmetry is then 
\begin{equation}\label{kappagauged}
\CP\, \Psi=\Psi~.
\end{equation}
Since the operator in \eqref{final_fermion_cusp} is rather involved, we will work out the fermionic spectrum in two specific limits, the cases $(\phi,\theta=0)$ (vanishing $q$) and $(\phi,\theta=0)$ ($p\to\infty$ keeping $k$ finite, such that $q/p=ik/\sqrt{1-k^2}$ and $b/p=1/\sqrt{1-k^2}$). In total analogy with the $\ads_5\times \sphere^5$ case, these are the two limits in which both the fermionic and the bosonic spectrum diagonalize, which makes it possible an exact evaluation of the corresponding partition functions.
%

\subsubsection{Partition function for $\theta=0$}
\label{app:theta0}

In this section we evaluate the partition function in the limit of vanishing $\theta$. This means to switch off the ``jump'' in $\cp^3$, setting $q$ in \eqref{def_pq} to zero. The string is  completely embedded in an $\ads_3$ subspace of $\ads_4$.
\paragraph{Bosonic fluctuations.}
In this limit the excitations $\zeta^7$ and $\zeta^8$ in (\ref{bosoniclagr})-(\ref{bosonicmasseslast}) decouple ($A=0$) and the bosonic modes reduce to six massless and two massive, one transverse and one longitudinal which can be rewritten in terms of the two-dimensional curvature  
\begin{equation}
M_{11}= 2\,, \qquad M_{ss} = M_{77} =0\,, \quad s=2,\dots,6\,, \qquad M_{88} = R^{(2)}+4\,,
\end{equation}
where $R^{(2)}$ is intended as the corresponding limit of (\ref{curvature}). 
Not surprisingly,  these are the same masses appearing in the $\ads_5\times S^5$ case at $\theta=0$~\cite{Drukker:2011za}. 
\paragraph{Fermionic fluctuations.}

For vanishing  $\theta$, the parameters in the Lagrangian \eqref{final_fermion_cusp} have the limits $\cos\zeta \rightarrow 1$ and $a\rightarrow a_0\equiv \frac{\sqrt{b^2}}{\sqrt{b^2+2}} \cosh\rho$. One finds then that Lagrangian $\mathcal{L}_F=  \bar\Psi D_F\Psi$  becomes
\begin{equation}\label{lagrfermtheta0}
\mathcal{L}_F=
i \bar\Psi\left\{ 
 \eta^{ab}\Gamma_a \p_b \CP 
 -{a_0(\sigma)\over 4} \left( \left( \Gamma_{69}-\Gamma_{58} + \Gamma_{47}\right)\Gamma_{11} +3 \Gamma_{0123}\right)\CP\right\}\Psi\, ,
\end{equation}
after rescaling the fermions by a factor $\sqrt{2 |a_0|}$. 
One can diagonalize the operator $\CP^T  \Delta_F \CP = - \CP^T  D^2_F \CP$ ($\CP$ commutes with $D_F$). 
One obtains thus  2 massless modes, 3 massive modes with $m_+$ and 3 with $m_-$ where $m_\pm$ are 
\begin{equation}
m^2_\pm =  a_0^2 \pm a'_0 =  \frac{1-k^2 \pm \sqrt{1-k^2} \dn\left(\sigma |k^2\right) \sn\left(\sigma | k^2\right)}{\cn \left(\sigma | k^2\right)^2}~.
\end{equation}
In the $\ads_5\times \sphere^5$ case at $\theta=0$~\cite{Drukker:2011za}, all fermions were massive with the same $m^2_\pm $ above.

\paragraph{Effective action.}

Due to the time-translational invariance the spectral problem is effectively one-dimensional. The resulting one-loop partition function can be then written as a single integral over  the Fourier-transformed $\tau$ variable $\partial_ \tau=-i\,\omega$ for the ratio of determinants of the operators corresponding to the bosonic and fermionic fluctuations above 
\begin{equation}
\label{effectiveaction_theta0}
\Gamma|_{\theta=0}= - \mathcal{T}  \int_{-\infty}^{+\infty} {d\omega\over 2 \pi} \log \frac{\det^{6/2} \O_F}{\det^{1/2} \O_1 \det^{4/2} \O_0 \det^{1/2} \O_2}\,.
\end{equation}
Above,  $\mathcal{T}=\int d\tau$ is the $\tau$-period, $\O_0 = -\p^2_\sigma+\omega^2$  and 
\begin{align}
  \label{operators_theta0}
  \O_1 &= -\p_\sigma^2 +\omega^2 +2 {(1-k^2)\over \cn^2 (\sigma)}\,, &
  &\O_2 = -\p_\sigma^2 +\omega^2 +2 {(1-k^2)\over \cn^2 (\sigma)} -2 k^2 \cn^2(\sigma)\,,
  \nn \\
  \O_F &=\O_+ = - \p_\sigma^2 +\omega^2 +m_+^2\,, &
  &\O_- (\sigma) = \O_+(\sigma +2\KK ) \,.
\end{align}
Above,  $\KK= \KK(k^2)$ and the last equivalence implies that $\det\O_F\equiv \det\O_+ =\det \O_-$. 
In~\cite{Drukker:2011za} these operators  could be put  into a single-gap Lam\'{e} form and thus evaluated analytically via the Gelfand-Yaglom method (see for example~\cite{Dunne:2007rt}). The latter is applied 
to a regularized initial value problem, where the Dirichlet boundary conditions (standard in this framework) are applied not at the extrema of the original $\sigma$-interval (\ref{range}), but considering $-\KK+\epsilon<\sigma<\KK-\epsilon$ with $\epsilon$ arbitrarily small. In Appendix~D of~\cite{Drukker:2011za} the reader can find the explicit form for the determinants, formulas (D.33)-(D.36), as well as a study of their small $\epsilon$ and large $\omega$  behavior. One can check that the redistribution of them within the integrand of (\ref{effectiveaction_theta0}) results in a small $\epsilon$, infrared divergent behavior $\sim\log(1/\epsilon^2)$ and in a large $\omega$, UV divergence $\sim\log(1/\omega^2)$. 
We could perform then a standard renormalization, consisting in an explicit subtraction of these divergences, for which the one-loop effective action would assume the form
\begin{equation}
\label{reg_effectiveaction_theta0_standard}
\Gamma^{\textrm{stand.}\,\textrm{renorm}}_{\theta=0} = -\frac{\mathcal T}{2} \lim_{\epsilon\rightarrow 0} \int_{-\infty}^{+\infty} \frac{d\omega}{2\pi}
 \log\Big[\, \frac{\epsilon^2\omega^2 \det^{6} \O^\epsilon_F} {\det \O^\epsilon_1 \det^{4} \O^\epsilon_0 \det \O^\epsilon_2}\,\Big]\,,
\end{equation}
where the superscript denotes the regularized determinants. However, in the $k\to0$ limit corresponding to the \emph{straight line}, this would result in a \emph{non-vanishing} finite  number which  can be explicitly evaluated to be equal to $1/8$. This is different from what happened in~\cite{Drukker:2011za}, where the standard renormalization was enough to reproduce, in the $k\to 0$ limit, the expected vanishing partition function corresponding to the BPS state. The difference, which extends to the  $\phi=0$ analysis considered   in Section \ref{app:phi=0}, is likely due to the presence of fermionic massless modes (\ref{lagrfermtheta0}) here and of ``light'' fermions (\ref{fermionsnewbis}) at $\phi=0$~\footnote{We have checked whether the finite contribution obtained in the $k\to0$ limit via the partition function with standard renormalization could result from the integrated effect of the contributions of massless free modes and lowest eigenvalues of the operators there appearing, much the same way the analysis done below in (\ref{E1intermediate})-(\ref{final2}) goes. This is not the case, since this contribution ends up to be proportional to $k$ and therefore vanishes in this limit.}.

To obtain the physically meaningful vanishing value in the $k\to0$ limit, we will take as natural reference the partition function corresponding to the $k\to0$ limit of our expressions. This is in fact the  partition function for a straight line in global coordinates in this $AdS_4\times \C P^3$ background. In the $k\to0$ limit, the $\varphi$ and $\vartheta$ coordinates have trivial values, and at the boundary ($\rho=\infty$) the straight line is described by the infinite time coordinate. The corresponding world-sheet in the AdS bulk, parametrized by coordinates whose range is now
\begin{equation}
-\frac{\pi}{2}<\sigma<\frac{\pi}{2}, \qquad -\infty<\tau\equiv t<+\infty~,
\end{equation}
is described by the $\rho$ coordinate (\ref{rhocn}), which  now extends all the way from the  boundary  to the center of AdS (where it vanishes) and then back to the boundary.  In the general case ($k\neq0$), the $\rho$ coordinate would have minimal value, an inversion point, at $\cosh^2\rho=\frac{1+b^2}{b^2}$. This inversion point coincides with the $AdS$ center $\rho=0$ when $k=0$  (being $b^2=(1-2 k^2)/k^2$). In the same limit the induced metric (\ref{ind-metric}) reads~\footnote{The classical description of the straight-line is equivalent to the one of~\cite{Drukker:2000ep,Kruczenski:2008zk,Chu:2009qt} in the $AdS_5\times S^5$ framework, the only differences being due to the fact that the analysis there performed is done in the Poincar\'e patch.}
\begin{equation}
{ds^2_\text{ind}}^{\!\!\!\!\!(k=0)}= \frac{1}{\cos^2\sigma}\big[{-}d\tau^2+d\sigma^2\big],
\label{ind-metric_k0}
\end{equation}
and the corresponding two-dimensional curvature is $R^{(2)}=-2$. The classical action (\ref{classical-action}) reduces in this limit to the regularized term
\begin{equation}
  {\mathcal{S}_\text{cl}}^{(k=0)}=\frac{T\sqrt{\bar{\lambda}}}{\pi}\,\sinh\rho_0\,.
  \label{classical-action-k0}
\end{equation}
At one loop one can perform a straightforward evaluation of the partition function, discovering not surprisingly that it can be written as in (\ref{effectiveaction_theta0}) with $k\to0$ limit of the fluctuations (\ref{operators_theta0}). We have checked that the corresponding determinants, evaluated via Gelfang-Yaglom method with Dirichlet boundary conditions using the regularized integral 
$-\frac{\pi}{2}+\epsilon<\sigma<\frac{\pi}{2}-\epsilon$, result in the $k\to0$ limit of the determinants appearing in (\ref{reg_effectiveaction_theta0_standard}), and 
amount to the same IR and UV divergences, which makes the resulting partition function a natural normalization for (\ref{reg_effectiveaction_theta0_standard}). 

\begin{figure}
  \centering
  \includegraphics[width=0.7\textwidth]{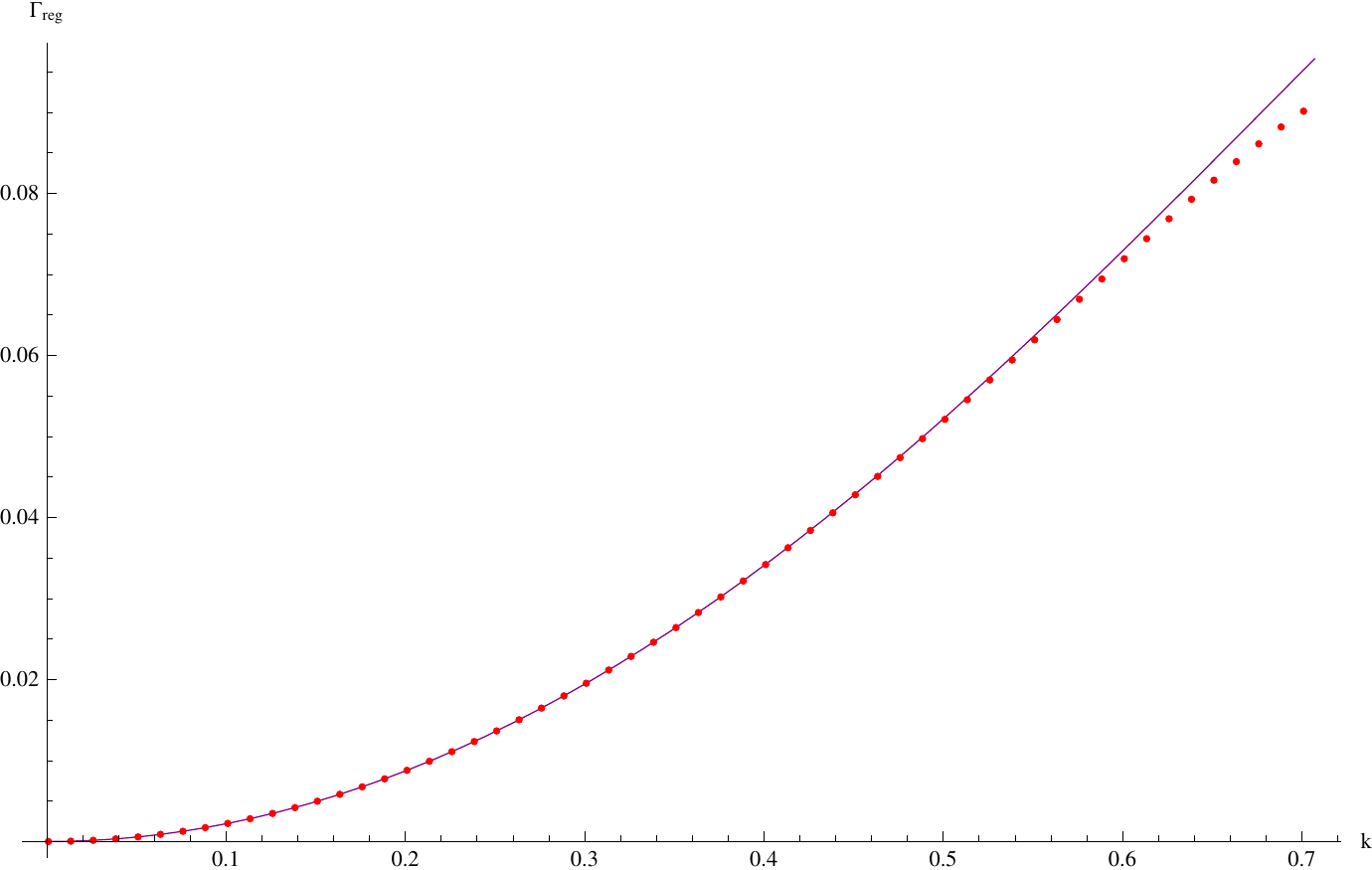}
  \caption{Plot of the partiton function for $\theta=0$. The straight purple line
    shows the small $k$ expansion~\ref{expansiontheta0} and the red dots show the
    result of numerically evaluating~\ref{reg_effectiveaction_theta0}.}
  \label{fig:plot-theta-0}
\end{figure}

The regularization prescription described above results then in the following one-loop effective action 
\begin{equation}
\label{reg_effectiveaction_theta0}
\Gamma^{\textrm{reg}}_{\theta=0} = -\frac{\mathcal T}{2}  \int_{-\infty}^{+\infty} \frac{d\omega}{2\pi}
 \log\Big[\, \frac{ \det^{6} \O'_F} {\det \O'_1 \det^{4} \O'_0 \det \O'_2}\,\Big]\,,
\end{equation}
where the prime refers to the normalization discussed before
\begin{equation}
\det\O_i'=\frac{\det \O^\epsilon_i}{{\det \O^\epsilon_i}|_{k=0}}~.
\end{equation}
From this exact one--loop contribution to the string partition function one derives the one--loop correction to the generalized cusp 
\begin{equation}\label{gammacusponeloopdef}
{\Gamma_{\textrm{cusp}}^{\textrm{ABJM}}}^{(1)}(\phi,0)=\frac{\Gamma^{\textrm{reg}}_{\theta=0}}{T}\,.
\end{equation}
The integral in (\ref{reg_effectiveaction_theta0}) can be evaluated numerically with very high precision. It is much more interesting, however, to expand its integrand in a power series for small $k$ and evaluate it analytically. 
Useful formulas for the expanded determinants can be found in~\cite{Drukker:2011za} (see section D.4 there) resulting in regular hyperbolic functions over which an integration can be always performed. We directly report here the results for the contributions to the regularized effective action (\ref{reg_effectiveaction_theta0}) up to order $k^6$ (where we omit the index $\theta=0$)
\begin{align}
 \Gamma_{\textrm{reg}} &= \sum_{i=1}^{\infty} \Gamma_{\textrm{reg} }^{(i)} k^{2 i} \,, \qquad
 \frac{\Gamma_{\rm reg}^{(0)} }{\mathcal T} =0\,,\qquad
 \frac{ \Gamma_{\rm reg}^{(2)}}{\mathcal T} = {7\over 32}\,,\no\\\label{results_theta0}
   \frac{ \Gamma_{\rm reg}^{(4)}}{\mathcal T} &=  {71 \over 512} - {9 \, \zeta(3)\over64 }\,, 
 \qquad
 \frac{ \Gamma_{\rm reg}^{(6)} }{\mathcal T} = \frac{289}{2048} - \frac{39}{512} \,\zeta(3)-\frac{45}{512} \,\zeta (5)\,.
\end{align}
It is useful to rewrite these results in terms of a $p\to\infty$ expansion, for which the connection to the straight-line expansion as an expansion in at small $\theta$ and $\phi$ can be made via (\ref{phithetapinfinity}).  The 1--loop correction (\ref{gammacusponeloopdef}) is then written as
\begin{align}
\label{expansiontheta0}
\frac{\Gamma_\text{reg}}{T}
&=\frac{1}{T}\Big[\Gamma_\text{reg}^{(0)}+k^2\Gamma_\text{reg}^{(2)}
+k^4\Gamma_\text{reg}^{(4)}+k^6\Gamma_\text{reg}^{(6)}
+\order(k^{8})\Big]\\
&=\frac{7}{32 p^2}-\frac{1}{p^4}\Big(\frac{153}{512} + \frac{9 \zeta (3)}{64}\Big)+
\frac{1}{p^6} \Big(\frac{1333}{2048} + \frac{321\zeta(3)}{512} - \frac{45\zeta(5)}{512}\Big)
+\order(p^{-8})\,,
\end{align}
where we have used (\ref{rescaling}) and (\ref{pq}) ($q=0$) from which in this limit follows
\begin{equation}
\mathcal{T}/T=\frac{1}{\sqrt{1-2 k^2}}
=\frac{(p^2+4)^{1/4}}{\sqrt{p}}\sim 1+p^{-2}-\textstyle{\frac{3}{2}} \,p^{-4}+\frac{7}{2}\,p^{-6}+O\,(p^{-8})\,.
\end{equation}
This analytic expansion is compared to a numerical evaluation of the partition
function~\eqref{reg_effectiveaction_theta0} in Figure~\ref{fig:plot-theta-0}, showing that its few leading terms give a very good approximation to the exact result.

\subsubsection{Partition function for $\phi=0$}
\label{app:phi=0}

In this section we evaluate the contribution of  fluctuations in the case in which minimal 
surface is within an $\ads_2\times \sphere^1$ subspace of 
$\ads_4\times\cp^3$. This corresponds to the limit $p\to\infty$ which keeps $k$ finite (namely $q/p=ik/\sqrt{1-k^2}$ and $b/p=1/\sqrt{1-k^2}$). 

\paragraph{Bosonic fluctuations.}
Also in this case the excitations $\zeta^7$ and $\zeta^8$ in (\ref{bosoniclagr})-(\ref{bosonicmasseslast}) decouple ($A=0$), the  fluctuation field $\zeta_8$ has the same action as $\zeta_1$,  and one notice, with respect to the corresponding $\ads_5\times \sphere^5$ case~\cite{Drukker:2011za}, the appearance of one heavy  ($M_{22}$) and four light ($M_{33}=\cdots=M_{66}$) modes in the transverse $\cp^3$ directions
\begin{align}\nonumber
M_{11}&=M_{88}=  2+\frac{k^2}{\sqrt{g}}\,,\qquad 
M_{22}=\frac{k^2}{\sqrt{g}} \,, \\
M_{77}&=R^{(2)}+2+\frac{k^2}{\sqrt{g}}\,, \qquad
M_{ss}=\frac{k^2}{4\,\sqrt{g}}\,, \quad s=3,\dotsc,6 \,.
\end{align}
Above, masses are written in terms of the  two-dimensional  scalar curvature  $R^{(2)}$ and the determinant of the world-sheet metric $\sqrt{g}$, again in the corresponding limits of (\ref{ind-metric}) and (\ref{curvature}). 

\paragraph{Fermionic fluctuations.}
In this limit the coefficient for the connection term $\Gamma_{137}$ in \eqref{final_fermion_cusp} becomes subleading in ${1\over p}$, thus we can neglect it, simplifying considerably the analysis. Then, the fermionic Lagrangian $\mathcal{L}_F = i  \bar\Psi D_F \Psi$ is
\begin{equation}
\mathcal{L}_F =  i \bar\Psi\left\{
\eta^{ab}\Gamma_a \p_b - {a_\infty(\sigma)\over 4} \left(\left( \Gamma_{69}- \Gamma_{58} + A (\sigma) \, \Gamma_{47}\right)\Gamma_{11}+3 A(\sigma) \, \Gamma_{0123}\right) \right\}\CP\, \Psi
\end{equation}
where $a_\infty = {\sqrt{1-k^2} \over \cn(\sigma| k^2) }$, $A = {\dn(\sigma| k^2) \over \sqrt{1-k^2}}$, and again we have rescaled the fermion by $\sqrt{2|a_\infty|}$. 
In order to simplify our analysis for the spectrum we use two further projectors  $P_\pm =\half \left( 1\pm \Gamma_{5689}\right)$, which naturally commute with $D_F$ and  with $\CP$. Thus, the eigenvalues will be the solutions of the characteristic equations for the operators
$P_+^T \CP^T \Delta_F \CP P_+$ and  $P_-^T \CP^T \Delta_F \CP P_-$.

\begin{figure}
  \centering
  \includegraphics[width=0.7\textwidth]{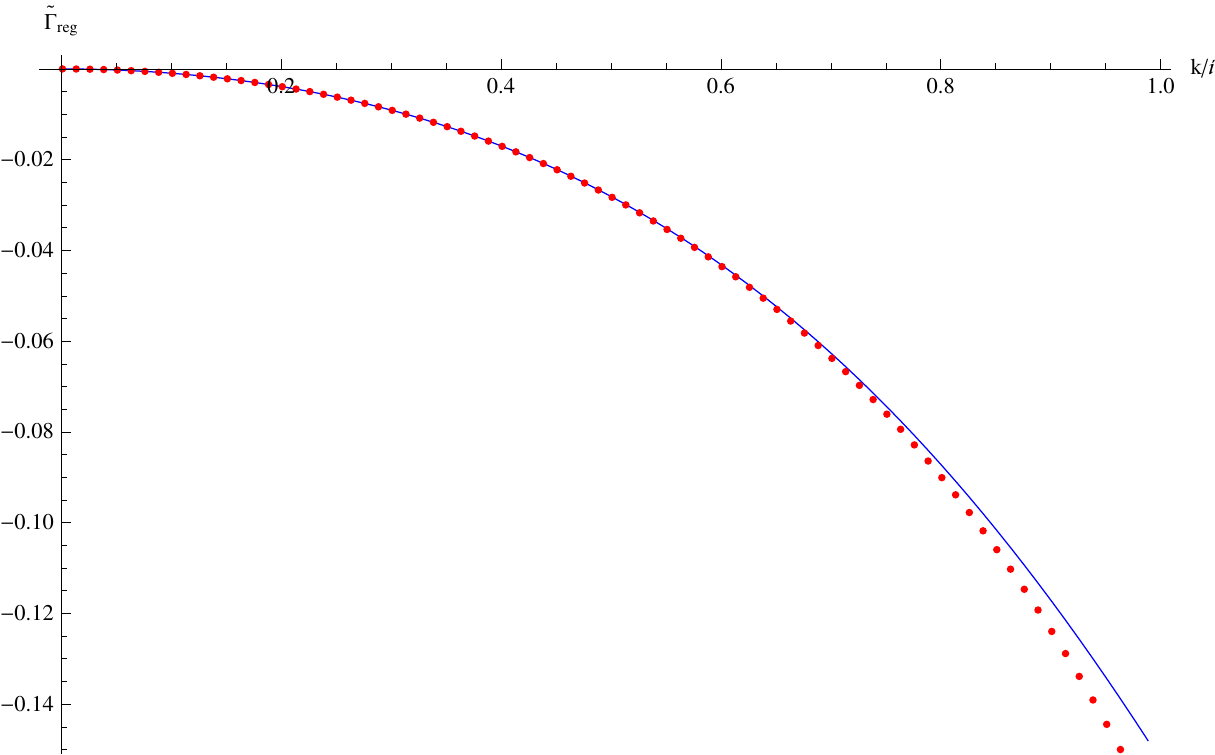}
  \caption{Plot of the partiton function for $\phi=0$. The straight purple line
    shows the small $k$ expansion~\ref{expansionphi0} and the red dots show the
    result of numerically evaluating~\ref{reg_effectiveaction_phi0}.}
  \label{fig:plot-phi-0}
\end{figure}

Explicitly, one obtains in the  first case (projection via $P_+$)   the masses 
\begin{equation}
\label{fermionsold}
m_\pm =  {1 \pm k^2 \sn(\sigma| k^2)\over 1\pm \sn(\sigma| k^2)}\,, 
\end{equation}
with multiplicity 2 each,  and for the projection via $P_-$
\begin{align}\label{fermionsnew}
  m_{1,2} &= {\pm 2 \pm 2 \sqrt{1-k^2} \dn(\sigma| k^2)+k^2 \left(\mp 1+ \sn(\sigma| k^2)\right)\over 4 \left(\pm 1+\sn(\sigma| k^2)\right)}\, \\
  \label{fermionsnewbis} 
  m_{3,4} &=  { \pm 2 \mp 2 \sqrt{1-k^2} \dn(\sigma| k^2)+k^2 \left(\mp1+ \sn(\sigma| k^2)\right)\over 4 \left(\pm1+\sn(\sigma| k^2)\right)}\, , 
\end{align}
The first two masses \eqref{fermionsold} are as in the corresponding $\phi=0$ case in $\ads_5\times \sphere^5$. Notice that the $k\to 0$ limits of  \eqref{fermionsold} and \eqref{fermionsnew} coincide. In that the masses (\ref{fermionsnewbis}) vanish in the  $k\to0 $ limit, one can consider the corresponding fermionic excitations as ``light'' in comparison to (\ref{fermionsold}).
\paragraph{Partition function.}
In writing down the explicit form of the partition function we can partially exploit the results of~\cite{Drukker:2011za}.
The determinants of the masses coinciding with their $\ads_5\times \sphere^5$ counterpart are written in formulas (E.15)-(E.18) there, where  $\det\widetilde{O}_0^\epsilon$  is now corresponding to the heavy transverse mode $M_{22}$, $\det{\widetilde{\mathcal{O}}}_1^\epsilon$ for $M_{11}=M_{88}$,  $\det{\widetilde{\mathcal{O}}}_2^\epsilon$ for $M_{77}$ and $\det{\widetilde{\mathcal{O}}}_F^\epsilon$ for the fermionic masses (\ref{fermionsold}).  The determinants for the light bosonic modes $M_{33}=\cdots=M_{66}$ are a slight modification of  $\det\widetilde{O}_0^\epsilon$, and read, at leading order in $1/\epsilon$ expansion,
\begin{equation}
\det\widetilde{O}_{0{\rm light}}^\epsilon\cong\frac{2\,\sinh \left(\KK\, \sqrt{k^2+4 \omega ^2}\right)}{\sqrt{k^2+4\omega ^2}}\,.
\end{equation}
About the fermionic  fluctuation operators with the newly appearing masses (\ref{fermionsnew})-(\ref{fermionsnewbis}), it is not surprising  that can be rewritten in Lam\'e form. Exploiting the same procedure as in~\cite{Drukker:2011za} one can check that the determinant which takes into account the total contributions of (\ref{fermionsnew}) and (\ref{fermionsnewbis}) read respectively, at leading order in $1/\epsilon$ expansion,
\begin{align}
\det\widetilde{\mathcal{O}}_{\textrm{F}_{1,2}}^\epsilon&\cong \frac{16 \,\sinh^2\big(\KK Z(\alpha_{F})\big) }{\epsilon^2\, (4 \omega ^2+1) (4 \omega ^2+1-k^2)}\,, \\\label{fermionlight}
\det\widetilde{\mathcal{O}}_{\textrm{F}_{3,4}}^\epsilon&\equiv \det\widetilde{\mathcal{O}}_{\textrm{Flight}}^\epsilon\cong\frac{\cosh ^2\big(\KK Z(\alpha_{F})\big) }{\omega ^2}\,, 
\end{align}
where 
\begin{equation}
\sn(\alpha_F|k^2)=\sqrt{\frac{4 \omega ^2+1}{k^2}}\,.
\end{equation}
As in the $\theta=0$ case discussed above, regularizing the resulting IR- and UV-divergent ratio of determinants multiplying it for $\epsilon^2\omega^2$ would result in a non physical $k\to0$ limit (we would obtain again a non-vanishing $1/8$ value for the partition function). We can proceed however adopting the same regularization prescription and write down the analog of (\ref{reg_effectiveaction_theta0}) reads here
\begin{equation}\label{reg_effectiveaction_phi0}
{\widetilde{\Gamma}_\text{reg}}|_{\phi=0}
=-\frac{\mathcal{T}}{2}\int_{-\infty}^{+\infty}\frac{d\omega}{2\pi}
\log\Big[\,\frac{\det^{4}{\widetilde{\mathcal{O}}}_F'\,\det\widetilde{\mathcal{O}}_{\textrm{F}_{1,2}}'\,\det{\widetilde{\mathcal{O}}}_{\rm Flight}'}
{\det^2{\widetilde{\mathcal{O}}}_1'\det{\widetilde{\mathcal{O}}}_2'\det{\widetilde{\mathcal{O}}}_{0}'\det^{4}{\widetilde{\mathcal{O}}}_{0{\rm light}}'}\,\Big]\,.
\end{equation}
where
\begin{equation}
{\widetilde{\mathcal{O}}}_i'=\frac{{\widetilde{\mathcal{O}}}_i^\epsilon}{{{\widetilde{\mathcal{O}}}_i^\epsilon}|_{k=0}}~.
\end{equation}
Once again, one can proceed with a small $k$ expansion of the determinants and work out explicit analytic results. We find in this case
\begin{align}
  \label{results_phi0}
  \frac{\widetilde{\Gamma}_{\rm reg}^{(0)}}{\mathcal{T}} &= 0\,, &
  \frac{ \widetilde{\Gamma}_{\rm reg}^{(2)} }{\mathcal{T}} &=  {3\over 32}\,,
  \\ \nn
  \widetilde{\Gamma}_{\rm reg}^{(4)} &= -\frac{1}{512}-\frac{\log (2)}{32}-\frac{3 \zeta (3)}{64}\,, &
  \widetilde{\Gamma}_{\rm reg}^{(6)} &=-\frac{15}{2048}-\frac{3 \log(2)}{128} -\frac{3 \zeta (3)}{128}+\frac{15 \zeta (5)}{512}\,.
\end{align}
In  a $p\to\infty$ expansion, one gets for the total one-loop correction 
\begin{align}\nonumber
\frac{  \widetilde{\Gamma}_\text{reg}}{T}
&=\frac{1}{T}\Big[  \widetilde{\Gamma}_\text{reg}^{(0)}+k^2  \widetilde{\Gamma}_\text{reg}^{(2)}
+k^4  \widetilde{\Gamma}_\text{reg}^{(4)}+k^6  \widetilde{\Gamma}_\text{reg}^{(6)}
+\order(k^{8})\Big]\\\label{expansionphi0}
&= -\frac{3}{32}\,\frac{q^2}{ p^2}-\frac{25 + 16\log 2 + 24\zeta(3)}{512}\frac{q^4}{ p^4}+\\\nonumber
&\qquad\qquad-\frac{3 (32 \zeta (3)+20 \zeta (5)+21+16 \log (2))}{2048}\frac{q^6} {p^6}+{\mathcal{O}}((q/p)^{8})\,,
\end{align}
where we have used (\ref{rescaling}) and (\ref{pq})  in this limit  
\begin{equation}
{\mathcal{T}}/T=\frac{1}{\sqrt{1-k^2}}\sim 1-\frac{q^2}{2 p^2}-\frac{q^4}{8\,p^4}
-\frac{q^6}{16\,p^4}-\frac{5\,q^8}{128\,p^8}+\mathcal{O}((q/p)^{10})\,.
\end{equation}
Figure~\ref{fig:plot-phi-0} shows a plot of the expansion
in~\eqref{expansionphi0} together with a numerical calculation of~\eqref{reg_effectiveaction_phi0}.

\subsection {Folded string}
\label{app:foldedABJM}

The classical parameters and conserved charges are the same as in the $\ads_5\times \sphere^5$ case, and we report them here for completeness.
In the folded string solution (\ref{solfolded}), $\rho$ varies from $0$ to its maximal  value $\rho_0$, which  is related to
  the useful parameter $\eta$  or   $k$   by 
\begin{equation}\label{kappaomega}
\coth^2 \rho_0 = \frac{\omega^2}{\kappa^2}\equiv 1+ \eta\equiv \frac{1}{k^2} \,,\qquad\k=\frac{2\,k}{\pi}\,\KK \,, \qquad \omega=\frac{2}{\pi}\,\KK \,,
\end{equation}
where the last two relations are obtained integrating the periodicity condition on the variable $\sigma$.
The  2-d metric induced by the solution (\ref{solfolded}) on the $(\tau,\s)$ cylinder  and its curvature are 
\begin{equation} \label{curv}
  g_{ab}=\r'^2(\s)\,\eta_{ab} \,, \qquad
  R^{(2)}=-\frac{\partial^2_\s\log\r'^2}{\r'^2}=-2+\frac{2\,\k^2\,\omega^2}{\r'^4}\,.
\end{equation}
The two conserved momenta conjugate to $t$ and $\phi$ are 
the classical energy and the spin 
\begin{equation}\label{ES}
  \frac{ E_0}{\sqrt{\lambda}}=\,\k\int_0^{2\pi}\frac{d\s}{2\pi}\cosh^2\r\equiv \,\E_0 \,, \qquad
  \frac{S}{\sqrt\lambda}=\,\omega\int_0^{2\pi}\frac{d\s}{2\pi}\sinh^2\r\equiv \,\S
\end{equation}
which in terms of the complete elliptic integrals  $\KK=\KK(k^2)$ and $\EE=\EE(k^2)$ read
\begin{equation}\label{Eclassic}
  \mathcal{E}_0 = \frac{2}{\pi}\,\frac{k}{1-k^2}\,\EE \,, \qquad
  \mathcal{S} = \frac{2}{\pi}\,\Big(\frac{1}{1-k^2}\,\EE-\KK \Big) \,.
\end{equation}
Finding the energy in terms of the spin (namely solving for $k$, or $\eta$,  in terms of $\S$ 
and then substituting it into the expression for  the energy $\E$) can be
 done in the two interesting limits of large spin (or long string  limit: $\r_0\to\infty$, i.e. $\eta\to0$ or $k\to 1$) 
\begin{equation}
\mathcal{E}_0= \S+\frac{\log (8\pi\S) -1}{\pi} + \frac{\log (8\pi\S) -1}{2 \pi^2 \S} + \dotsb  \,, \qquad
\S\gg1 \,,  \label{Ecl} 
\end{equation}
where the leading $\log S$  term is governed by the so-called ``scaling function'' (cusp anomaly), and small spin (or short string limit:  $\r_0\to0$, i.e. $\eta\to\infty$ or  $k\to0$). 
\begin{equation}
 \E_0=\sqrt{2\,\S}\Big(1 +\frac{3}{8}\,\S + \dotsb \Big) \,, \qquad
\S\ll1 . 
\label{Ec}
\end{equation}
which results in the usual flat-space Regge relation~\cite{Gubser:2002tv,Frolov:2002av}.

\bigskip

About \emph{fluctuations} over the solution (\ref{solfolded}), in the case of the bosons they are obtained as a simple truncation of the $\ads_5\times \sphere^5$ case, as explained in the main text. The analysis for fermions was already carried out in~\cite{McLoughlin:2008ms,McLoughlin:2008he}, and we report here just the final steps.
After suitable global rotation and rescaling, and again reorganizing the spinors via the projector operator (\ref{def_P}) one obtains\footnote{We have slightly different notation with respect to~\cite{McLoughlin:2008ms,McLoughlin:2008he}, due to a different labeling and ordering for the $\cp^3$ coordinates. }
\begin{equation}
\label{L_final_final}
\CL_F =
i \Big\{ 
 \bar\Psi \Gamma_a\p^a\mathcal P\Psi
+\frac{e^\phi k}{8}\rho' 
\left[ \bar\Psi \left( \left( -\Gamma_{47}+\Gamma_{58}-\Gamma_{69}\right)\Gamma_{11}-3\Gamma_{0123}\right)\mathcal P \Psi
\right]\Big\}\,,
\end{equation}
where again the linear combination~$\Psi=\Psi^1+\Psi^2$ has been used. Using the $\kappa$-symmetry gauge fixing (\ref{kappagauged}), the fermionic part of the quadratic Lagrangian can can be written in the form
$\CL_F = \bar\Psi\,{\rm D}_F \Psi$,
where
\begin{equation}\label{fermionlagrangeanfoldedABJM}
  {\rm D}_F=i \Big(\Gamma_a\p^a  
  +\,\rho'\, \widetilde{\Gamma} \,\Big) \,,\qquad
  \widetilde{\Gamma}=  \frac{1}{4}\left(- \Gamma_{47}+\Gamma_{58}-\Gamma_{69}\right)\Gamma_{11}-\frac{3}{4}\,\Gamma_{0123} \,.
\end{equation}
since now $e^\phi k=2$. Using $\log\det({\rm D}_F)=\frac{1}{2}\log\det({\rm D}_F)^2$ , and squaring the corresponding Dirac operator in (\ref{fermionlagrangeanfoldedABJM}), one obtains that the fermionic contribution to the 2-d  effective action is governed by  the following differential operator
\begin{equation}
 \Delta_F=({\rm D}_F)^2=-\partial_a\partial^a-\r''\,\Gamma_1 \widetilde{\Gamma} - \rho '^2\, (\widetilde{\Gamma})^2\,.
\end{equation}
This operator can in fact be further diagonalized and one ends up with 8 effective fermionic degrees of freedom, of which two are massless and the remaining 6 have masses $\rho'^2\pm\rho''$.
  
We report here an alternative  expression for the one-loop energy  (\ref{final}) useful in the short string limit, which follows from separating the contributions of  the  massless modes of   $\O_0$  (i.e.  $\Om^2$) and 
the  lowest analytically known eigenvalues of the operators  $\O_\theta,\O_\phi$  and $\O_{\psi}$ (see the detailed analysis of~\cite{Beccaria:2010ry}, and Table 1 there). One gets 
\begin{equation}\label{E1intermediate}
 E_1=-\frac{{1}}{4\pi\k }\int_{-\infty}^{\infty}
 \mathrm{d}\Om\,\left[\log \frac{(\det'\mathcal{O}_\psi)^6}{    \det' \mathcal{O}_\phi\   
 \det'\mathcal{O}_{\theta} \   (\det'\mathcal{O}_0)^4
   } + h(\Om)\right]
\end{equation}
where
\begin{gather}\label{detprime}
{\det}'\mathcal{O}_{\theta} \equiv\frac{\det\mathcal{O}_{\theta}}{\k^2+\Om^2}\,, \qquad
{\det}'\mathcal{O}_{\phi,\psi,0} \equiv \frac{\det\mathcal{O}_{\phi,\psi,0}}{\Om^2} \,, \\
 h(\Om)=\log(\Om^2)-\log(\Om^2+\k^2) \,.  \label{hj}
\end{gather}
Using that  $\int_{-\infty} ^\infty d\Om\  h(\Om^2)=- 2 \,\pi\,\k $, the one-loop correction to the energy (\ref{final}) takes the form 
\begin{equation}\label{final2}
 E_1=\frac{1}{2}-\frac{1}{4\pi\kappa}\int_{-\infty}^{\infty}\mathrm{d}\Om \ 
 \,\log \frac{(\det'\mathcal{O}_\psi)^6}{\det'\mathcal{O}_\phi \ \det'\mathcal{O}_{\theta} \   (\det'\mathcal{O}_0)^4} \,. 
\end{equation}

One makes use of the general expansion of the determinants, see~\cite{Beccaria:2010ry}, to get
the short string expansion of (\ref{final2}) 
\begin{align}\no
 E_1 &=\frac{1}{2}-\frac{1}{4\pi\kappa}\int_{-\infty}^{\infty}\mathrm{d}\Om \ \log \frac{(\det'\mathcal{O}_\psi)^6}{\det'\mathcal{O}_{\theta} \det'\mathcal{O}_\phi\,(\det'\mathcal{O}_0)^4}\\
 &=\frac{1}{2}+ \frac{1}{\k} \Big[
 \textstyle{\big(1-3\,\log2\big)}\eta^{-1}+\big(-\frac{21}{32}+\frac{9}{8}\log2+\frac{9}{32}\,\zeta(3)\big)\eta^{-2}+\O\,(\eta^{-3}) \Big]\,.
\end{align}
Substituting $\frac{1}{\kappa} =  \sqrt{ \eta} \, \big[1 + \frac{1}{ 4  }\eta^{-1}  +   \order(\eta^{-3}) \big]$
and the expansion of $\eta$ in terms of the spin (\ref{ES}), one writes (\ref{smallspin}).

\section{Spinning strings in $\ads_3 \times \sphere^3\times \sphere^3\times \sphere^1$}
\label{app:AdS3CFT2}

In this appendix we consider spinning strings in Type IIB string theory on
$\ads_3 \times \sphere^3\times \sphere^3\times \sphere^1$. We first recollect
some basic facts about this background. In section~\ref{sec:spinning_ads3_cft2}
we write down a classical string solution carrying angular momentum in $\ads$ as
well as on both three-spheres. In the following section we derive the bosonic
and fermionic quadratic fluctuation Lagrangians around this background. These
results are the basis of the one-loop analysis in~\ref{sec:intro_ads3_cft2},
where we expand te exact one-loop partition function around the long and short
string limits in the case where only the angular momentum in the $\ads$ part of
the background is non-trivial. To complement the result in the main text we also
consider spinning strings in the scaling limit, where all angular momenta are
non-zero. These results can be found in section~\ref{sec:ads3-scaling-limit}. In
section~\ref{sec:cusp-in-ads3} we finally consider the embedding of the
generalized cusp solution from~\cite{Drukker:2011za} into $\ads_3 \times
\sphere^3\times \sphere^3\times \sphere^1$.

The metric of the $\ads_3 \times \sphere^3 \times \sphere^3 \times \sphere^1$
background is given by
\begin{gather}
  \label{metric_ads3}
  ds^2 = R^2 \left( ds^2_{\ads}+{1\over \alpha} ds^2_{\sphere^3_+}+{1\over 1-\alpha} ds^2_{\sphere^3_-}\right) + dU^2\,,
  \\ \nn
  ds^2_{\ads} = -\cosh^2\rho\,dt^2 +d \rho^2 +\sinh^2\rho\,d\phi^2\,,\qquad
  ds^2_{\sphere^3_\pm} = d\beta^2_\pm +\cos^2\beta_\pm\left( d\gamma_\pm^2 + \cos^2\gamma_\pm\,d\varphi_\pm^2\right)\,.
\end{gather}
The background additionally contains a non-trivial Ramond-Ramond three-form
\begin{equation}
  e^{\phi} F = \vol(\ads_3)  + \frac{1}{\alpha} \vol(\sphere^3_+)  + \frac{1}{1-\alpha} \vol(\sphere^3_-) \,.
\end{equation}
The supergravity equations of motion relate the radii $R_\pm$ of the two
three-spheres to the AdS radius $R$ via a ``triangle equality''~\cite{Gauntlett:1998kc}
\begin{equation}
  \frac{1}{R^2} = \frac{1}{R_+^2} + \frac{1}{R_-^2} \,.
\end{equation}
A useful parametrization of the relation above is\footnote{%
  The same parameter $\alpha$ also appears in the exceptional Lie superalgebra
  $\alg{d}(2,1;\alpha)$, two copies of which form the super-isometries of
  $\ads_3 \times \sphere^3 \times \sphere^3$. Notice that the choice
  \eqref{alpha}, for which $\alpha$ must lie between zero and one, corresponds
  to the choice of the real form of $\alg{d}(2,1;\alpha)$, which is the one of
  interest to us~\cite{Babichenko:2009dk}.%
} %
\begin{equation}\label{alpha}
  \alpha = \frac{R^2}{R_+^2 } \,, \qquad
  1-\alpha = \frac{R^2}{R_-^2} \,.
\end{equation}
Special values of $\alpha$ of interest to us are $\alpha\to 1$, arising when the
radii of AdS and the one of the two spheres become equal and the other sphere
decompactifies, and $\alpha\to\frac{1}{2}$, corresponding to the two spheres
having the same size. The $\alpha\to 1$ limit is of particular importance,
since the resulting setting should be Type IIB string theory on $\ads_3\times
\sphere^3\times \torus^4$, as suggested by the coset analysis
in~\cite{Babichenko:2009dk}.\footnote{%
  Concerning the isometries of the background, when $\alpha\to 1$ the
  $\alg{d}(2, 1; \alpha)^2$ algebra turns into $\mathfrak{psu}(1, 1|2)^2$, which
  is the algebra preserved by $\ads_3 \times \sphere^3 \times \torus^4$. From
  the fact that $\alg{d}(2, 1; \alpha)$ is isomorphic to $\alg{d}(2, 1;
  1-\alpha)$, and from (\ref{alpha}), it is clear that the only difference
  between the $\alpha\to0$ and $\alpha\to1$ is in choice of which of the two
  three-spheres becomes decompactified.%
} %
\subsection{Spinning string}
\label{sec:spinning_ads3_cft2}

A classical string carrying three spins in $\ads_3\times \sphere^3\times \sphere^3\times
\sphere^1$ is obtained from the ansatz
\begin{equation}
  \label{classical_spinning}
  \begin{aligned}
    t &= \kappa \tau\,, &
    \phi &= \omega\tau\,, &
    \rho(\sigma) &= \rho(\sigma+2\pi)\,,
    \\
    \varphi_+ &= \nu_+ \tau\,,&
    \varphi_- &= \nu_- \tau\,, &
    \gamma_\pm &=\beta_\pm= U=0\,.
  \end{aligned}
\end{equation}
The equation of motion  coincides with the one for its $\ads_5\times \sphere^5$ counterpart~\cite{Frolov:2002av}
\begin{equation} \label{eom-AdS3}
  \rho'' = \cosh \rho\,\sinh\rho\,(\kappa^2 -\omega^2).
\end{equation}
The Virasoro constraint (implying (\ref{eom-AdS3}) for $\rho'\neq0$) reads
\begin{equation}
  \rho'^2 = \kappa^2\cosh^2 \rho -\omega^2 \sinh^2 \rho -\nu^2 \,,
\end{equation}
where 
\begin{equation}
  \label{def_nu}
  \nu^2 \equiv \frac{\nu_+^2}{\alpha} + \frac{\nu^2_-}{1-\alpha} \,.
\end{equation}
In order to further quantify the distribution of the angular momentum between the two spheres we introduce
the parametrization 
\begin{equation}
  \label{def-delta-parametrization}
  \nu_+=\sqrt{\alpha\,\delta}\,\nu \,, \qquad
  \nu_-=\sqrt{(1-\alpha)\,(1-\delta)}\,\nu \,,
\end{equation}
where $0 \le \delta \le 1$.

The above solution carries the Noether charges~\cite{Gubser:2002tv,Frolov:2002av}
\begin{align}
  \mathcal{E} \equiv \tfrac{1}{\sqrt{\lambda}} E &= \int_0^{2\pi} \frac{d\sigma}{2\pi} \, \kappa \cosh^2 \rho \,, \\
  \mathcal{S} \equiv \tfrac{1}{\sqrt{ \lambda}} S &= \int_0^{2\pi} \frac{d\sigma}{2\pi} \, \omega \sinh^2 \rho \,, \\
  \mathcal{J}_+ \equiv \tfrac{1}{\sqrt{ \lambda}} J_+ &= \int_0^{2\pi} \frac{d\sigma}{2\pi} \, \frac{\nu_+}{\alpha} = \sqrt{\frac{\delta}{\alpha}}\,\nu \,, \\  
  \mathcal{J}_- \equiv \tfrac{1}{\sqrt{ \lambda}} J_- &= \int_0^{2\pi} \frac{d\sigma}{2\pi} \, \frac{\nu_-}{1-\alpha} = \sqrt{\frac{1-\delta}{1-\alpha}}\,\nu \,.  
\end{align}
It is convenient to perform a rotation in the $\varphi_\pm$
directions by introducing
\begin{equation}
  \varphi = \sqrt{\frac{\delta}{\alpha}} \varphi_+ + \sqrt{\frac{1-\delta}{1-\alpha}} \varphi_- \,, \qquad
  \psi = - \sqrt{\frac{1-\delta}{\alpha}} \varphi_+ + \sqrt{\frac{\delta}{1-\alpha}} \varphi_- \,.
\end{equation}
The classical solution in these directions then takes the form
\begin{equation}
  \varphi = \nu\tau \,, \qquad
  \psi = 0 \,.
\end{equation}
We also introduce the corresponding angular momenta
\begin{gather}
  \mathcal{J} \equiv \tfrac{1}{\sqrt{\lambda}} J = \sqrt{\alpha\delta}\,\mathcal{J}_+ + \sqrt{(1-\alpha)(1-\delta)}\,\mathcal{J}_- = \nu \equiv \sqrt{\lambda}\,\mathcal{J} \,, \\
  \mathcal{K} \equiv \tfrac{1}{\sqrt{\lambda}} K = -\sqrt{\alpha(1-\delta)}\,J_+ + \sqrt{(1-\alpha)\delta}\,J_- = 0 \,.
\end{gather}
The classical energy $\mathcal{E}$ of the string can now be obtained as a
function of the charges $\mathcal{S}$ and $\mathcal{J}$ and can be expressed in
terms of elliptic integrals. We will not write down this expression here, but
just note that it takes the same form as for the spinning string in $\ads_5
\times \sphere^5$~\cite{Gubser:2002tv,Frolov:2002av,Frolov:2006qe}.

\subsection{Bosonic fluctuations}
\label{sec:ads3-bos-fluct}

The leading quantum correction to the energy of the solution can be evaluated by
expanding the action to quadratic order in fluctuations near the classical
solution (\ref{classical_spinning}) and computing the corresponding partition
function expressed in terms of determinants of the quadratic fluctuation 
operators. The fluctuations are introduced by
\begin{align}
  \label{bosonexpansion1}
  \nn
  t &= \kappa \tau +\frac{\tilde t}{\lambda^{1/4}} \,,
  &
  \rho &= \rho(\sigma)+\frac{\tilde \rho}{\lambda^{1/4}} \,,
  &
  \phi &= \omega \tau+\frac{\tilde \phi}{\lambda^{1/4}} \,,
  &
  \varphi &= \nu\tau + \frac{\tau\varphi}{\lambda^{1/4}} \,,
  \\ \nn
  \beta_+ &= \frac{\sqrt{\alpha}} {\lambda^{1/4}}\,\tilde\beta_+ \,,
  &
  \gamma_+ &=\frac{\sqrt{\alpha}} {\lambda^{1/4}}\,\tilde\gamma_+ \,,
  &
 \beta_- &= \frac{\sqrt{1-\alpha}} {\lambda^{1/4}}\,\tilde\beta_- \,,
  &
  \gamma_- &=\frac{\sqrt{1-\alpha}} {\lambda^{1/4}}\,\tilde\gamma_- \,,
  \\
  &&
  \psi &= \frac{\tilde \psi}{\lambda^{1/4}} \,,
  &
  U &=\frac{\tilde U}{\lambda^{1/4}} \,,
\end{align}
where, for convenience, we have rescaled the fluctuations $\tilde{\beta}_\pm$ and $\tilde{\gamma}_\pm$ by constant factors. 
Using the standard static gauge  ($\tilde t=0$ and $\tilde\rho=0$), and in units where the radius of $\ads_3$ is set to one, the bosonic fluctuation Lagrangian reads
\begin{align}
  2 \mathcal{L}_{2B} &=
  \sinh^2\bar\rho \left( 1 + \frac{\omega^2\sinh^2\bar\rho}{(\bar\rho')^2} \right) \partial^a \tilde\phi \, \partial_a \tilde\phi
  + \left( 1 + \frac{\nu^2}{(\bar{\rho}')^2} \right) \partial^a \tilde\varphi \, \partial_a \tilde\varphi
  + \frac{2 \omega \nu \sinh^2\bar\rho}{(\bar\rho')^2} \partial^a \tilde\phi \, \partial_a \tilde\varphi 
  \nn \\ &\phantom{{}=}
  + \partial^a \tilde\beta_+ \, \partial_a \tilde\beta_+ + \partial^a \tilde\gamma_+ \, \partial_a \tilde\gamma_+
  + \partial^a \tilde\beta_- \, \partial_a \tilde\beta_- + \partial^a \tilde\gamma_- \, \partial_a \tilde\gamma_-
  + \partial^a \tilde{\psi} \, \partial_a \tilde{\psi}
  + \partial^a \tilde{U} \, \partial_a \tilde{U} 
  \nn \\ &\phantom{{}=} \label{LAdS3S3S3}
  + \nu_+^2 ( \tilde\beta_+^2 + \tilde\gamma_+^2 )
  + \nu_-^2 ( \tilde\beta_-^2 + \tilde\gamma_-^2 )
  \,.
\end{align}
The first line of the equation above coincides as expected with the coupled part
of the (static gauge) bosonic fluctuation Lagrangian for a folded string in
$\ads_3\times \sphere^1$ embedded into $\ads_5\times
\sphere^5$~\cite{Iwashita:2010tg}.\footnote{%
  The ``transverse'' fluctuations, not coupled, obviously differ. In
  $\ads_3\times \sphere^3\times \sphere^3\times \sphere^1$ we have two massless
  fluctuations, one in $\sphere^3 \times \sphere^3$ and one in the $\sphere^1$,
  as well as two massive fluctuations in the first $\sphere^3$ (constant mass
  $\nu^2_+$) and two in the second $\sphere^3$ (constant mass $\nu^2_-$). In the
  $\ads_5\times \sphere^5$ case one gets four transverse fluctuations in $S_5$
  with constant masses $\nu^2$ and two in $\ads_5$, with non-constant masses
  $2\rho'^2+\nu^2$.%
} %
We note that the $\alpha$- and $\delta$-dependence only resides in the masses of
the trivial (non-coupled) fields $\beta_\pm$ and $\gamma_\pm$. For
$\alpha=\delta=1/2$ these fields have mass $\nu/2$ which coincide with the
masses of the light transversal fluctuations around a spinning string in $\ads_4
\times \cp^3$~\cite{McLoughlin:2008ms,Alday:2008ut}. For $\alpha=\delta=0$ and
$\alpha=\delta=1$, on the other hand, two of the excitations become massless and
the other two have mass $\nu$, like the fluctuations on the sphere in $\ads_5
\times \sphere^5$~\cite{Frolov:2002av}. For a further discussion on this limit see
section~\ref{AdS3-S3-T4}.

\subsection{Fermionic fluctuations}
\label{app:fermionsADS3}

The quadratic part of the Green-Schwarz action for fermions reads~\cite{Cvetic:1999zs}
\begin{equation}\label{LGS}
  \mathcal{L}_{\textrm{GS}}= i \left( \sqrt{-h} h^{ab} \delta^{IJ} -\epsilon^{ab} \sigma_3^{IJ}\right) \bar\theta^I \rho_a  D_b^{JK}  \theta^{K} \,,
\end{equation}
where the super covariant derivative is
\begin{equation}\label{covder}
  D_b^{JK}  \theta^{K}= \delta^{JK}\left( \p_b+\frac{1}{4} \omega_M^{AB} \p_b X^M \Gamma_{AB}\right) \theta^K+\frac{1}{24} F_{MNP} \Gamma^{MNP}  \rho_b \, \sigma_1^{JK} \theta^K\,.
\end{equation}
Above,  the coupling to the RR three-form flux is written explicitly as
\begin{equation}\label{flux_ads3}
 F_{MNP} \Gamma^{MNP} = 6 \left( \Gamma^{012}+\sqrt{\alpha} \Gamma^{345}+\sqrt{1-\alpha} \Gamma^{678}\right)\,.
\end{equation}
For the classical solution \eqref{classical_spinning}, the  2-d projections of $\Gamma$-matrices  $\rho_a =\p_a X^M E_M^A \Gamma_A$ are
\begin{equation}
  \rho_0 = \kappa \cosh\rho \Gamma_0+\omega \sinh\rho \Gamma_2 +{\nu_+\over \sqrt \alpha} \Gamma_5+{\nu_-\over \sqrt{1-\alpha}} \Gamma_8\,,\qquad
  \rho_1 = \rho' \,\Gamma_1 \,.
\end{equation}
To put the GS action (\ref{LGS}) in a suitable form for Ò2-d fermion in curved 2-d spaceÓ, one can perform a local rotation of the spinors ($\theta^I = S \Psi^I$) using the  combination of  boosts  
\begin{equation}\label{boost}
  S = \left( \cosh{\frac{\xi}{2}}+ \sinh{\frac{\xi}{2}}\Gamma_{02}\right) \left(  \cosh{\frac{\zeta}{2}}+\sinh{\frac{\zeta}{2}}\Gamma_0 \bar\Gamma\right)\,,
\end{equation}
where
\begin{equation}
  \bar\Gamma \equiv \sqrt{\delta} \, \Gamma_5 + \sqrt{1-\delta} \, \Gamma_8
\end{equation}
and the parameters in (\ref{boost}) are defined by
\begin{equation}
 \cosh{{\xi}} = \frac{\kappa \cosh\rho}{\sqrt{\rho'^2+\nu^2}}\,, 
  \qquad
  \cosh\zeta=  \frac{\sqrt{\rho'^2+\nu^2}}{\rho'}\,.
 \end{equation}
It then immediately follows that the set of $\sigma$-dependent 10-d Dirac matrices are transformed into ten constant Dirac matrices
$  \tau_a\equiv S^{-1} \rho_a S$, $\tau_0 = \rho'\Gamma_0$, $\tau_1= \rho' \Gamma_1$, 
and
\begin{gather}
  S^{-1} \p_0 S =\p_0\,,
  \quad
  S^{-1} \p_1 S =  
 \p_1 + \frac{\kappa\omega}{2} \frac{\Gamma_{02}}{\sqrt{\rho'^2+\nu^2}}
  + \frac{\kappa \omega \nu}{2} \frac{\Gamma_2\bar\Gamma}{\rho'^2+\nu^2}   
  - \frac{\rho''}{2\rho'} \frac{\nu\,\Gamma_0\bar\Gamma}{\sqrt{\rho'^2+\nu^2}}\,,
  \nn \\ 
  S^{-1} \frac{1}{4} \p_0 X^M \omega_M^{BC} \Gamma_{BC} S  = \frac{\rho''}{2 \rho' }\Gamma_{01} - \frac{\kappa \omega}{2 \sqrt{\rho'^2+\nu^2}} \Gamma_{12} 
  + \frac{\rho''}{2 \rho' \sqrt{\rho'^2+\nu^2}} \Gamma_1\bar\Gamma\,.
\end{gather}
Rotating similarly the flux term in (\ref{covder}), choosing the gauge-fixing  $\Psi^1=\Psi^2 \equiv \Psi$ and rescaling the fermions by a factor $\sqrt{2|\rho'|}$, the complete Lagrangian \eqref{LGS} reads
\begin{align}
\mathcal L_{\textrm{GS}} =
 i \bar\Psi\Big\{
&\Gamma^a\p_a - \frac{\kappa\, \omega\,\nu}{2 (\rho'^2+\nu^2)} \Gamma_{12}\bar{\Gamma} 
+ \frac{\rho'}{2} \Big( 
\Gamma^{012} - \sqrt{\alpha}  \Gamma^{345} - \sqrt{1-\alpha}  \Gamma^{678}
\Big)
\nn \\& \label{LGS-AdS3S3S3}
+ 
\frac{\sqrt{\rho'^2+\nu^2} - \rho'}{2}
\Big(
\Gamma^{012}
- \left(\sqrt{\alpha\delta} \Gamma^{34} + \sqrt{(1-\alpha)(1-\delta)} \Gamma^{67}\right) \bar{\Gamma}
\Big) 
\Big\} \Psi\,,
\end{align}
Note that with the gauge choice $\Psi^1=\Psi^2 \equiv \Psi$ only the symmetric
part of the Lagrangian \eqref{LGS} contributes.

\subsection{The scaling limit in the long string approximation}
\label{sec:ads3-scaling-limit}

The spectrum of quadratic fluctuations around the spinning string significantly
simplifies in the limit where we take $\kappa \gg 1$ with $u = \nu/\kappa$
fixed~\cite{Frolov:2006qe}. From the Virasoro constraint we then find
\begin{equation}
  \rho'(\sigma) = \sqrt{\kappa^2 - \nu^2} \,, \qquad
  \omega = \kappa.
\end{equation}
We will restrict our analysis to the case $\delta=\alpha$.\footnote{%
  We also note the $\nu \to \kappa$ limit, where the string becomes point-like and
  moves along a null geodesic~\cite{Babichenko:2009dk}. In order to write the masses of
  the fluctuations in a compact form we introduce $\alpha \equiv \sin^2 u$ and
  $\delta \equiv \sin^2 v$. The bosons then have masses $0$, $\kappa \sin
  u\,\sin v$, $\kappa \cos u\,\cos v$ and $1$, while the fermionic masses are
  $\kappa \sin^2\frac{u\pm v}{2}$ and $\kappa \cos^2\frac{u\pm v}{2}$, with all
  modes doubly degenerate. In particular, for $v = u$ the point-like string is
  supersymmetric, and the spectrum reduces to the BMN-like spectrum
  of~\cite{Babichenko:2009dk}.%
} %
The bosonic Lagrangian~\eqref{LAdS3S3S3} then describes eight fluctuation modes
with the frequencies
\begin{gather}
  \omega^B_{1,2}(n) = n \,, \qquad
  \omega^B_{3,4}(n) = \sqrt{n^2 + \alpha^2\nu^2} \,, \qquad
  \omega^B_{5,6}(n) = \sqrt{n^2 + (1-\alpha)^2\nu^2} \,, \\
  \label{scaling-limit-ads-boson-disp-rel}
  \omega^B_{7,8}(n) = \sqrt{n^2 + 2\kappa^2 \mp 2\sqrt{n^2\nu^2 + \kappa^4}} \,.
\end{gather}
From the quadratic Green-Schwarz action in~\eqref{LGS-AdS3S3S3} we find four fermionic modes with frequencies
\begin{equation}
  \omega^F_{1,2}(n) = \pm\frac{\nu}{2} + n \,, \qquad
  \omega^F_{3,4}(n) = \pm\frac{\nu}{2} + \sqrt{n^2 + \kappa^2} \,.
\end{equation}
The frequencies of the four other modes are given by solutions to the quartic equations~\footnote{We thank A. Tseytlin for pointing out a misprint in the sign  in front of $ \big(\tfrac{1}{2}-\alpha\big)^2\nu^2$  in the l.h.s. of \eqref{quartic-equation} in the previous version of this paper.}
\begin{equation}
  \label{quartic-equation}
  \bigg((\omega^F_i)^2 - n^2 - \big(\tfrac{1}{2}-\alpha\big)^2\nu^2 \bigg)^2 
  = \bigg(\kappa^2 \Big(\omega^F_i \pm \big(\tfrac{1}{2}-\alpha\big)\nu\Big)^2 - (\kappa^2-\nu^2) n^2 \bigg) \,.
\end{equation}
These equations are straightforward to solve, but the general solutions are
quite involved. Here we will consider the special case of $\alpha=1/2$, where
the two three-sphere have the same radius. We then find the frequencies
\begin{equation}
  \omega^F_{5,6}(n) = \sqrt{n^2 + \frac{\kappa^2}{2} + \sqrt{n^2\nu^2 + \frac{\kappa^4}{4}}} \,, \qquad
  \omega^F_{7,8}(n) = \sqrt{n^2 + \frac{\kappa^2}{2} - \sqrt{n^2\nu^2 + \frac{\kappa^4}{4}}} \,.
\end{equation}

The one-loop correction to the string energy can be found by summing over the
quadratic fluctuations~\cite{Frolov:2002av}. Hence we need to calculate the sum
\begin{equation}
  E_1^{\alpha=1/2} = \frac{1}{2\kappa} \sum_{n=-\infty}^{\infty} \Bigg[
  \sum_{i=1}^{8} \omega_i^B(n) - \sum_{i=1}^{8} \omega_i^F(n)
  \Bigg]\,.
\end{equation}
For $\kappa \gg 1$ we can replace the summation with integration. The various
terms in this integral are very similar to what was found in the one-loop
analysis of the spinning string in $\ads_4 \times \cp^3$ in~\cite{McLoughlin:2008ms},
and can be calculated in a similar fashion. The final result for the one-loop
correction to the string energy then takes the form
\begin{equation}
  \label{eq:E1-1/2}
  \begin{split}
    E_1^{\alpha=1/2} =\frac{\nu}{2u} \bigg(
    -\big(1-u^2\big) &{}+ \sqrt{1-u^2} - u^2\log u \\ &{}- (2-u^2)\log\Big(1+\sqrt{1-u^2}\Big) - 2\big(1-u^2\big)\log 2
    \bigg) \,.
  \end{split}
\end{equation}
This expression is very similar to the $\ads_4 \times \cp^3$ result
of~\cite{McLoughlin:2008ms}. Comparing the two energies we have
\begin{equation}
  \label{eq:scaling-AdS4-vs-12}
  E_1^{\alpha=1/2} - E_1^{\ads_4} = \frac{\nu}{2u} \bigg( u^2 \log u + (2-u^2) \log\sqrt{2-u^2} \bigg).
\end{equation}

Taking the $u \to 0$ limit in~\eqref{eq:E1-1/2} and including also the
tree-level energy we obtain
\begin{equation}
  E - S = f(\lambda) \log S \,, \qquad
  f(\lambda) = \frac{\sqrt{\lambda}}{\pi} - \frac{2\log 2}{\pi} \,,
\end{equation}
where we used that $\kappa = (\log S)/\pi$. This result for the scaling function
$f(\lambda)$ for $\ads_3$ is identical to what we found in
section~\ref{sec:ads3-expansion}.

In~\cite{OhlssonSax:2011ms} it was proposed that the spectrum in the $\grSL(2)$
sector for $\alpha = 1/2$ is described Bethe equations
\begin{equation}
  \label{eq:BA-sl2}
  \left(\frac{x_i^+}{x_i^-}\right)^L =
  - \prod_{\substack{k = 1\\k \neq i}}^{K} \frac{x_i^- - x_k^+}{x_i^+ - x_k^-} \frac{1 - \frac{1}{x_i^+ x_k^-}}{1 - \frac{1}{x_i^- x_k^+}} \sigma^2(x_i,x_k) \,. 
\end{equation}
The form of these equations is exactly the same as for the corresponding sector
in ABJM~\cite{Gromov:2008qe}. This similarity is only superficial since
here is no a priori reason for the dressing phase factor $\sigma$ to take the
same form in the two models.\footnote{%
  As discussed in~\cite{Babichenko:2009dk,OhlssonSax:2011ms} the leading strong coupling
  behavior of $\sigma$ should take the same form as in $\ads_5 \times
  \sphere^5$~\cite{Arutyunov:2004vx}.%
} %
However, let us assume that also the one-loop correction to the phase
agrees. We can then use the results from~\cite{Gromov:2008qe} for the energy of
the spinning string, which give
\begin{equation}
  E - S = f(h) \log S + \dotsb,
\end{equation}
with
\begin{equation}
  \label{eq:h-expansion-12}
  f(h) = 2 h - \frac{3\log 2}{2\pi} + \order(1/h) \,.
\end{equation}
Hence we find that the above result from the Bethe ansatz agrees with our string
theory calculation provided
\begin{equation}
  h = \frac{\sqrt{\lambda}}{2\pi} - \frac{\log 2}{4\pi} + \order(1/\sqrt{\lambda}) \,, \qquad \alpha = 1/2 \,.
\end{equation}
This correction to $h(\lambda)$ is similar to what is found in the analysis of
$\ads_4 \times  \cp^3$ in~\cite{McLoughlin:2008he}. However, there has been some
controversy over this result. In~\cite{Gromov:2008fy}, the same calculation
was performed using the algebraic curve~\cite{Gromov:2008bz}. The result of
this calculation was that the one-loop correction to $h(\lambda)$ was
trivial. The difference between the two calculations is in the regularization of
the UV-divergences. In the world-sheet calculations in~\cite{McLoughlin:2008ms,McLoughlin:2008he}
(see also~\cite{Alday:2008ut,Krishnan:2008zs}) the same cutoff was used for all
modes. In the algebraic curve analysis, on the other hand, it is more natural to
split the modes into ``light'' and ``heavy'' modes, and impose different cutoffs
in the two sectors so that $\Lambda_{\text{heavy}} =
2\Lambda_{\text{light}}$. It would be interesting to redo the above calculation
using the algebraic curve in~\cite{Babichenko:2009dk} to see if there is
a similar ambiguity also in $\ads_3$.

It is straightforward to repeat this analysis for $\alpha=1$. The spectrum is
now even simpler. There are four massless bosons, two bosons with mass $\nu$ and
two bosons in $\ads$ with dispersion relations given
by~\eqref{scaling-limit-ads-boson-disp-rel}. Furthermore, there are four massless
fermions as well as four fermions with mass $\kappa$. This spectrum gives the
one-loop energy
\begin{equation}
  E_1^{\alpha=0} =\frac{\nu}{u} \bigg( \!\!
  -\big(1-u^2\big) + \sqrt{1-u^2} - u^2\log u - (2-u^2)\log\Big(1+\sqrt{1-u^2}\Big)
  \! \bigg) \,.
\end{equation}
In this case it is natural to compare our result to that in $\ads_5 \times
\sphere^5$~\cite{Frolov:2006qe}
\begin{equation}
  \label{eq:scaling-AdS5-vs-1}
  E_1^{\alpha=0} - E_1^{\ads_5} = \frac{\nu}{u} \bigg( u^2 \log u + (2-u^2) \log\sqrt{2-u^2} \bigg).
\end{equation}
Note that the right-hand sides of equations~\eqref{eq:scaling-AdS4-vs-12}
and~\eqref{eq:scaling-AdS5-vs-1} differ only by an overall factor two.

In~\cite{Babichenko:2009dk} a set of Bethe equations for the spectrum at $\alpha=1$
were given. The relevant equation for the case at hand is
\begin{equation}
 \label{eq:BA-sl2-T4}
 \left(\frac{x_i^+}{x_i^-}\right)^L =
 \prod_{\substack{k = 1\\k \neq i}}^{K} \frac{x_i^- - x_k^+}{x_i^+ - x_k^-} \frac{1 - \frac{1}{x_i^+ x_k^-}}{1 - \frac{1}{x_i^- x_k^+}} \sigma^2(x_i,x_k) \,.
\end{equation}
Note that this equation differs from~\eqref{eq:BA-sl2} by a minus sign on the
right-hand side. This makes~\eqref{eq:BA-sl2-T4} identical to the Bethe
equation for the $\grSL(2)$ sector of $\ads_5 \times \sphere^5$. If we
again assume that the dressing phase $\sigma$ is the same in the two models at
next-to-leading order at strong coupling, we find the scaling function~\cite{Casteill:2007ct}
\begin{equation}
  f(h) = 2h - \frac{3\log2}{\pi} + \order(1/h).
\end{equation}
Hence we now need to have
\begin{equation}
  \label{eq:h-expansion-1}
  h = \frac{\sqrt{\lambda}}{2\pi} + \frac{\log 2}{2\pi} + \order(1/\sqrt{\lambda}) \,, \qquad \alpha = 1 \,.
\end{equation}
Comparing~\eqref{eq:h-expansion-12} and~\eqref{eq:h-expansion-1} we note that
the one-loop correction to $h(\lambda)$ predicted by our calculation is
different in the two cases. In particular the correction is negative at
$\alpha=1/2$ and positive at $\alpha=0$.

\subsection{Generalized cusp in $\ads_3 \times \sphere^3 \times \sphere^3 \times \sphere^1$}
\label{sec:cusp-in-ads3}

For completeness we will in this section report on the action of quadratic fluctuations around the same solution considered in section~\ref{sec:cusp}. Such solution lives in $\ads_3\times \sphere^1$, hence it can be straightforwardly embedded in $\ads_3 \times
\sphere^3 \times \sphere^3 \times \sphere^1$. 

The
classical solution is given by
\begin{equation}
  \begin{aligned}
    t &= \tau \,, &
    \phi &= \sigma \,, &
    \rho &= \rho(\sigma) \,, \qquad
    U = 0 \,, \\
    \beta_+ &= 0 \,, &
    \gamma_+ &= 0 \,, &
    \varphi_+ &= \sqrt{\alpha\delta} \varphi(\sigma) \,, \\
    \beta_- &= 0 \,, &
    \gamma_- &= 0 \,, &
    \varphi_- &= \sqrt{(1-\alpha)(1-\delta)} \varphi(\sigma) \,.
  \end{aligned}
\end{equation}
The parameter $\delta$ describes the embedding of the non-trivial circle of the
solution into $\sphere^3 \times \sphere^3$. It is again useful to define the
coordinates
\begin{equation}
  \varphi = \sqrt{\frac{\delta}{\alpha}} \varphi_+ + \sqrt{\frac{1-\delta}{1-\alpha}} \varphi_- \,, \qquad
  \psi = - \sqrt{\frac{1-\delta}{\alpha}} \varphi_+ + \sqrt{\frac{\delta}{1-\alpha}} \varphi_- \,,
\end{equation}
so that the classical solution takes the form $\varphi = \varphi(\sigma)$, $\psi = 0$.

Introducing fluctuations in the same way as in~\eqref{bosonexpansion1} and using
the same static gauge as was used for the $\ads_4 \times  \cp^3$ calculation we
find that the action for the bosonic fluctuations can be written in terms of
eight fluctuations $\zeta_P$, $P = 1,\dotsc, 8$, as
\begin{equation}
  \mathcal{L}_{2B} = \frac{1}{2} \sqrt{g}  \Big( g^{ab} \partial_a \zeta_P \partial_b \zeta_Q + A (\zeta_8 \partial_\sigma \zeta_7 - \zeta_7 \partial_\sigma \zeta_8) + M_{PQ} \zeta_P \zeta_Q \Big) \,,
\end{equation}
where $g$ is the induced world sheet metric. The coefficients $M_{77}$, $M_{78}
= M_{87}$, $M_{88}$ and $A$ take the same form as in $\ads_4 \times  \cp^3$,
see~\eqref{bosonicmasseslast}. The rest of the mass matrix $M$ is diagonal with the
entries
\begin{equation}
  \label{eq:cusp-ads3-bos}
  \begin{aligned}
    M_{ii} &= 0 \,, && i = 1, 2 \,, \\
    M_{ss} &= \frac{b^4 - b^2p^2 - p^2}{b^2p^2\cosh^2\rho} \alpha\delta \,, && s = 3,4\,, \\
    M_{ss} &= \frac{b^4 - b^2p^2 - p^2}{b^2p^2\cosh^2\rho} (1-\alpha)(1-\delta) \,, && s = 5,6\,.
  \end{aligned}
\end{equation}
The four massive modes have masses that are proportional to the massive modes on
the sphere in $\ads_5 \times \sphere^5$, with $\alpha$- and $\delta$-dependent
coefficients.

After fixing $\kappa$-gauge and performing a rotation in spinor space, the
quadratic Green-Schwarz action takes the form
\begin{align}
  \label{eq:cusp-ads3-ferm}
  \mathcal{L}_{\textrm{GS}} = i \sqrt{-h} \bar{\Psi} \bigg[
 &h^{ab} \tau_a \partial_b - \frac{1}{2} \frac{\xi' - \cosh\rho}{a} \sin\zeta\,\Gamma_{12}\tilde{\Gamma} 
  + \frac{1}{2} \Big( \Gamma^{012} - \sqrt{\alpha}\,\Gamma^{345} - \sqrt{1-\alpha}\,\Gamma^{678} \Big) 
  \nonumber \\
 &\quad- \sin^2\frac{\zeta}{2} \Big( \Gamma^{012} - \big( \sqrt{\alpha\delta}\,\Gamma^{34} + \sqrt{(1-\alpha)(1-\delta)}\,\Gamma^{67} \big) \tilde{\Gamma} \Big)
  \bigg] \Psi \,,
\end{align}
where the rotation parameters take the same form as in $\ads_4$
(see~\eqref{xizetacusp}).

In the limit $q \to 0$ limit considered for the $\ads_4$ case in
section~\ref{sec:lines} the Lagrangian~\eqref{eq:cusp-ads3-ferm} describes four
massless fermions and four fermions of unit mass. We also get seven massless
bosons and one boson of mass two. Hence the spectrum we find in this limit
agrees with the results of~\cite{Drukker:2000ep} for all values of $\alpha$ and $\delta$.


\section{ Fluctuations about the folded string rotating in $\ads_3\times \sphere^1$}
\label{app:ads5xs5}

We revisit here the analysis of the bosonic and fermionic spectrum of fluctuations above the closed string solution rotating in $\ads_3\times \sphere^1$~\cite{Frolov:2002av}, showing that to  the coupled system of bosonic modes~\cite{Frolov:2002av,Iwashita:2010tg} corresponds a non trivial fermionic mass matrix.

\bigskip

In the evaluation of the \emph{fermionic spectrum}, we follow the steps and notation of~\cite{Frolov:2002av}. We apply to the fermionic Lagrangian
\begin{equation}
\mathcal{L}_{F}=i(\sqrt{-g}\,g^{ab}\,\delta^{IJ}-\epsilon^{ab}\, s^{IJ})\,
\Big[\bar\theta^I\,\rho_a\mathcal{D}_b\theta^J-\frac{i}{2}\epsilon^{JK}\,\bar\theta^I\rho_a\Gamma_*\,\rho_b\,\theta^K\Big] \,,
\end{equation}
evaluated on the solution $t=\kappa\,\tau,~\rho=\rho(\sigma),~\phi=\omega\,\tau,~\varphi=\nu\,\tau$, two local boosts in the $(0,2)$ and $(0,9)$ planes
\begin{equation}
S=\exp{\Big(\frac{\alpha_1}{2}\Gamma_0\,\Gamma_2\Big)}\,\exp{\Big(\frac{\alpha_2}{2}\Gamma_0\,\Gamma_9\Big)}\,.
\end{equation}
where the indices $(0, 1, 2, 9)$ are used to label the $t, \rho,\phi,\varphi$ directions in the tangent space. 
After the field redefinition $\theta^I\to\psi^I\equiv S^{-1}\theta^I$ and choosing the $\psi^1=\psi^2$ $\kappa$-symmetry gauge  one gets  ($\sqrt{-g}=\rho'^2$)
\begin{equation}
\mathcal{L}_{F}= 2i\,\sqrt{-g}\,\bar\psi\,\Big[\,g^{ab}\,\tau_a\mathcal{D}'_b-\sqrt{\frac{\rho'^2+\nu^2}{\rho'^2}}\Gamma_{234}\,\Big] \,\psi,
\end{equation}
where  $(\tau_0,\tau_1)=\rho'\,(\Gamma_0,\Gamma_1)$ and
 $\mathcal{D}'_j=\partial_j+S^{-1}\,\partial_j\,S+\frac{1}{4}\Omega^{ab}_j\,S^{-1}\Gamma_{ab}S$.
Since
\begin{equation}
\eta^{ab}\Gamma_a\,\mathcal{D}'_b =-\Gamma^0\,\Big(\partial_\tau+\frac{\rho''}{2\rho'}\Gamma_0\Gamma_1\,\Big)+\Gamma^1\,\Big(\,\partial_\sigma+\frac{\nu\,\kappa\,\omega}{2(\rho'^2+\nu^2)}\Gamma_2\Gamma_9\,\Big) \,,
\end{equation}
and using the two-dimensional connections $\nabla_\tau=\partial_\tau+\frac{\rho''}{2\,\rho'}\,\Gamma_0\Gamma_1, ~\nabla_\sigma=\partial_\sigma$, the Lagrangian reads
\begin{equation}
\mathcal{L}_F=2\,i\,\bar\psi\Big[\tau^a\,\nabla_a-\sqrt{\rho'^2+\nu^2}\,\rho'\,\Gamma_{234}+\frac{\nu\,\kappa\,\omega\,\,\rho'}{2(\rho'^2+\nu^2)}\,\Gamma_{129}\,\Big]\,\psi\,.
\end{equation}
Weyl-rescaling the fermions ($\psi\to\rho'^{-1/2}\psi$) the fermionic fluctuation Lagrangian is finally
\begin{equation}\label{fermop}
\mathcal{L}_F = 2i\,\bar\psi\,D_F\,\psi,\qquad D_F=\Gamma^a\,\partial_a+a(\sigma)\,\Gamma_{234}+b(\sigma)\,\Gamma_{129} \,,
\end{equation}
with
\begin{equation}\label{fermop2}
a(\sigma)=-\sqrt{\rho'^2+\nu^2}, \qquad b(\sigma)=\frac{\nu\,\kappa\,\omega}{2(\rho'^2+\nu^2)}\,.
\end{equation}
With respect to (5.41) in~\cite{Frolov:2002av}, the additional connection-related term $b(\sigma)\,\Gamma_{129}$ is present\footnote{We thank A. Tseytlin for discussing with us this new feature, and thank him and Y.\ Iwashita for double-checking the correctness of (\ref{fermop})-(\ref{fermop2}) via the Pohlmeyer-reduced theory action expanded near the folded string string solution with orbital momentum $J$~\cite{Iwashita:2010tg}.}. \\
The fermionic operator (\ref{fermop}) can be then rewritten as
\begin{equation}\label{fermopbis}
D_F=\Gamma^a\,\text{D}_a+a(\sigma)\,\Gamma_{234}\,, \qquad{\rm D}_a=\partial_a+A_a\,, \quad \text{with} \quad A_0=0\,,\quad A_1=b(\sigma)\,\Gamma_{29} \,,
\end{equation}
and one can get rid of the non-trivial connection term $A_1$  performing a local rotation
\begin{equation}
T=\exp\Big(\frac{\beta}{2}\Gamma_2\Gamma_9\Big)\,.
\end{equation}
The rotated fermionic operator $\widetilde D_F$ defined by $\mathcal{L}_F=2\,\bar\Psi\,\widetilde D_F\Psi$, where  $\Psi=T^{-1}\psi$, reads then
\begin{equation}\label{DFtilde}
\widetilde D_F:=T^{-1}\,D_F\,T=i\,\Big[\Gamma^a\partial_a+a(\sigma)\,\Gamma_{34}\,(\cos\beta\,\Gamma_{2}+\sin\beta\,\Gamma_{9}\,)+\Big(b(\sigma)+\frac{\beta'}{2}\Big)\,\Gamma_{129}\,\Big],
\end{equation}
and the contribution $\Gamma_{129}$ disappears by requiring $\beta'=-2 \,b(\sigma)$, namely\footnote{This is the same as  (C.27) in~\cite{Drukker:2011za}.}
\begin{equation}\label{betarotationf}
\beta'=-\frac{\nu\,\kappa\,\omega}{\rho'^2+\nu^2} \qquad \text{or} \qquad \beta(\sigma)=-i\frac{\nu\kappa\omega}{\kappa\sqrt{\rho'^2+\nu^2}}\,\Pi\Big(\frac{\kappa^2-\omega^2}{\kappa^2},i\,\rho,\frac{\kappa^2-\omega^2}{\kappa^2-\nu^2}\Big)\,.
\end{equation}
Squaring now $\widetilde D_F$, one obtains ($'$ represents the derivative with respect to $\sigma$)
\begin{align}\label{DFtilde2}
\widetilde D_F^2&=-\partial_a\partial^a+a^2(\sigma)-\Big(a(\sigma)\,\cos\beta(\sigma)\Big)'\,\Gamma_{1234}-\Big(a(\sigma)\,\sin\beta(\sigma)\Big)'\,\Gamma_{1349}\\
&=-\partial_a\partial^a+{\bf X}_F \,,
\end{align} 
The highly non-trivial mass matrix ${\bf X}_F$ has a simple trace
\begin{equation}
  \tr \mathbf{X}_F=16\,a^2(\sigma)= 16\,(\rho'^2+\nu^2) \,,
\end{equation}
which is exactly as in~\cite{Frolov:2002av}, where 8 two-dimensional fermions were found with mass $\rho'^2+\nu^2$.
Thus, the analysis of the logarithmic divergences coming from the fermionic part of the action\footnote{Recall that they do not cancel against the ones corresponding to the 2-d bosonic theory, leaving a divergence proportional to the two-dimensional curvature.} coincides with the one in~\cite{Frolov:2002av}.

Next, let us consider the operator (\ref{fermop}) in the limit of $\rho'$ constant, when  
\begin{equation}
\kappa\approx\omega \,,\qquad \rho'^2+\nu^2\approx\kappa^2 \,,  \qquad a\approx -\kappa \,, \qquad b\approx\frac{\nu}{2}\,.
\end{equation}
The operator has then constant coefficients
\begin{equation}
D_F|_{\rho'={\rm const}}=i\,\Big[\Gamma^a\,\partial_a+a\,\Gamma_{234}+b\,\Gamma_{129}\Big]
=i\,\Big[\Gamma^a\,\partial_a-\kappa\,\Gamma_{234}+\frac{\nu}{2}\,\Gamma_{129}\Big]~
\end{equation}
and the spectrum of characteristic frequencies is immediately computable.
This corresponds to 4+4 effective fermionic degrees of freedom with frequencies
\begin{equation}
\Omega_{F_{\pm}}=\sqrt{n^2 + \kappa^2} \pm \frac{\nu}{2} \,,\qquad n=0,\pm1,\pm2, \dotsc
\end{equation}
In the sum over all frequencies the shifts $\pm \frac{\nu}{2}$ cancel each other, the fermions behave  as 8 fermionic degrees of freedom with frequency $\sqrt{n^2+\k^2}$, as considered in~\cite{Frolov:2002av} and~\cite{Frolov:2006qe}.

\bigskip

The comparison with the \emph{bosonic spectrum} becomes more transparent when
performed in a static gauge where the fluctuations along the world--sheet
directions are set to zero. Therefore, expanding the bosonic action to quadratic
order in fluctuations near the classical background
\begin{equation}
  t=\kappa\,\tau+\frac{\tilde{t}}{\lambda^{1/4}} \,, \qquad
  \rho=\rho(\sigma)+\frac{\tilde{\rho}}{\lambda^{1/4}} \,, \qquad
  \phi=\omega\,\tau+\frac{\tilde{\phi}}{\lambda^{1/4}} \,, \qquad
  \varphi=\nu\,\tau+\frac{\tilde{\varphi}}{\lambda^{1/4}}
\end{equation}
one should not consider fluctuations of $\rho$ ($\tilde\rho=0$) and should  project out one direction parallel to the world--sheet in the $\tau$ direction. 
One then chooses two linear combinations 
of $\tilde t$, $\tilde\phi$ and $\tilde\varphi$ 
which are normal to the world--sheet and freeze  the third direction, which is tangential\footnote{The same procedure was adopted in~\cite{Drukker:2011za}, where indeed this symmetry between the bosonic and fermionic spectrum has been first observed.}.
For the normal directions one then chooses
\begin{align}
\zeta_7&=\frac{\kappa\, \sinh ^2\rho\,  \left(\rho'^2+\nu^2-\omega ^2\right)\,\tilde\phi\,-\omega  \,\cosh^2\rho\,\left(\rho'^2+\nu ^2-\kappa ^2\right) \,\tilde t}{\sqrt{\nu ^2+\rho'^2} \sqrt{\rho'^2+\nu ^2-\omega ^2}
   \sqrt{\rho'^2+\nu ^2-\kappa^2}}
\\
\zeta_8&=\frac{\kappa \, \nu\,  \cosh ^2\rho\, \tilde t + \left(\rho'^2+\nu ^2\right)\,\tilde \theta-\nu \, \omega  \sinh ^2\rho\,\tilde\phi}{\rho'\,\sqrt{\rho'^2+\nu ^2}}\,,
\end{align}
where $\rho$, $\rho'$ are evaluated for the classical solution, and  $\zeta_7$ and $\zeta_8$ are unit normalized so to have
canonical kinetic terms. 
The resulting action takes the form
\begin{equation}
  \mathcal{L}_{B}=\frac{1}{2}\sqrt{g}\Big[g^{ab}\,\partial_a \zeta_P\,\partial_b \zeta_P
  +A(\zeta_8\partial_\sigma\zeta_7-\zeta_7\partial_\sigma\zeta_8)
  +M_{PQ}\zeta_P\zeta_Q\Big],
  \qquad
  P,Q=1,\cdots,8
\end{equation}
with (recall $\sqrt{g}=\rho'^2$)
\begin{equation}
  \begin{gathered}
    \frac{1}{2}\,\sqrt{g}\,A =\frac{\kappa\,\nu\,\omega}{\rho'^2+\nu^2} \,, \qquad
    M_{78} =M_{87}=\frac{\kappa \, \nu \, \omega\, \rho''\, (\rho'^2-\nu^2)}{\rho'^3 \,(\nu ^2+\rho'^2)^2}\,,
    \\
    M_{77} =2+\frac{2 \kappa ^2 \omega ^2}{\rho'^2 \,(\nu ^2+\rho'^2)}+\frac{\nu ^2}{\rho'^2} \left(\frac{\kappa ^2 \omega ^2}{\left(\nu ^2+\rho'^2\right)^2}+1\right)
    \,,
    \\
    M_{88}=\frac{\nu ^2}{\rho'^2}-\frac{\kappa ^2 \omega ^2 \left(\nu ^2+2 \rho'^2\right)}{\rho'^2 \left(\nu ^2+\rho'^2\right)^2}-\frac{2 \left(\kappa ^2-\nu
        ^2\right) \left(\nu ^2-\omega ^2\right)}{\rho'^4}\,,
    \\
    M_{ii} =\frac{2\,\rho'^2+\nu^2}{\rho'^2} \,, \quad  i=1,2\,, \qquad 
    M_{ss}=\frac{\nu^2}{\rho'^2}\,, \quad s=3,4,5,6\,.
  \end{gathered}
\end{equation}
A field redefinition should relate the expressions here obtained with the ones following by setting, in static gauge, $\tilde\rho=\tilde t=0$.
In the $\nu\to 0 $ limit $A=0$ and the the first-order terms do not anymore contribute,  $M_{77}|_{\nu=0} = 2 \rho'^2 + \frac{2\kappa^2 \omega^2}{\rho'^2} $ and the massless mode $M_{88}|_{\nu\to0}=0$ as  expected.

The first order terms can be eliminated by the $\sigma$-dependent rotation 
\begin{equation}
\begin{pmatrix}\zeta_7\\\zeta_8\end{pmatrix}
\to\begin{pmatrix}\cos\beta&\sin\beta\\-\sin\beta&\cos\beta\end{pmatrix}
\begin{pmatrix}\zeta_7\\\zeta_8\end{pmatrix},
\end{equation}
where $\beta(\sigma)$ solves the equation 
\begin{equation}\label{betarotation}
\beta' =-\frac{1}{2}\sqrt{g}A\,=\,-\frac{\kappa  \nu  \omega }{\nu ^2+\rho'^2}\,.
\end{equation}
This also shifts the masses ($M_{PQ}\to \widetilde{M}_{PQ}$) and one ends with the Lagrangian
\begin{equation}
\mathcal{L}_{B}=\frac{1}{2}\sqrt{g}\Big[g^{ab}\,\partial_a \zeta_P\,\partial_b \zeta_P
+\widetilde{M}_{PQ}\zeta_P\zeta_Q\Big],
\end{equation}
where
\begin{align}\label{bosonshift}
M_{77}&\to
\widetilde{M}_{77}=\frac{1}{2}\left(M_{77}+M_{88}+(M_{77}-M_{88})\cos2\beta-M_{78}\sin2\beta-\frac{\sqrt{g}}{2}A^2\right),
\\
M_{88}&\to\widetilde{M}_{88}=
\frac{1}{2}\left(M_{77}+M_{88}-(M_{77}-M_{88})\cos2\beta+M_{78}\sin2\beta-\frac{\sqrt{g}}{2}A^2\right),
\\
M_{78}&\to\widetilde{M}_{78}=
\frac{1}{2}\big(M_{78}\cos2\beta+(M_{77}-M_{88})\sin2\beta\big)
\end{align}
and the remaining masses stay the same.
It is important to notice that the same rotation (\ref{betarotationf}) that eliminates connection-related terms in the  fermionic Lagrangian is also the one, (\ref{betarotation}), which cancel the first-order terms in the bosonic fluctuation Lagrangian. 
%


\bibliographystyle{nb}
\bibliography{references}

\end{document}